\documentclass[12pt]{iopart}
\pdfoutput=1
\usepackage{graphicx}
\usepackage[position=t,singlelinecheck=off]{subfig}
\usepackage{caption}
\usepackage{amssymb}
\usepackage[english]{babel}
\usepackage{psfrag}
\usepackage{hyperref}
\newcommand{\bq}{\begin{equation}}
\newcommand{\eq}{\end{equation}}
\newcommand{\ba}{\begin{eqnarray}}
\newcommand{\ea}{\end{eqnarray}}
\usepackage{color}

\begin{document}

\title{A simple mechanism for controlling the onset and arrest of collective oscillations in genetic circuits}
\author{D Labavi\'{c} and H Meyer-Ortmanns}
\address{School of Engineering and Science, Jacobs University Bremen, P.O.Box 750561, D-28725 Bremen, Germany}
\eads{\mailto{h.ortmanns@jacobs-university.de}}

\begin{abstract}
We study a system of dynamical units, each of which shows excitable or oscillatory behavior, depending on the choice of parameters. When we couple these units with repressive bonds, we can control the duration of collective oscillations for an intermediate period between collective fixed-point behavior. The control mechanism works by monotonically increasing a single bifurcation parameter. Both the onset and arrest of oscillations are due to bifurcations. Depending on the coupling strength, the network topology and the tuning speed, our numerical simulations reveal a rich dynamics out-of-equilibrium with multiple inherent time scales, long transients towards the stationary states and interesting transient patterns like self-organized pacemakers. Zooming into the transition regime, we pursue the arrest of oscillations along the arms of spirals. We point out possible relations to the genetic network of segmentation clocks.
 \end{abstract}


\maketitle

\section{Introduction}
\label{intro_sec}
The motivation for our study of coupled dynamical units, each of which showing excitable or oscillatory behavior, comes from the segmentation clock. The segmentation clock stands for an oscillating multicellular genetic network that controls the sequential subdivision of the elongating vertebrate embryonic body
axis into morphological somites (these are regularly sized cell clusters). The morphogenetic rhythm of the somitogenesis is based on repeated waves of oscillatory gene expression sweeping through the tissue. Usually it is the clock and wavefront mechanism \cite{1976} that is supposed to translate  the oscillations in time of a clock into a periodic pattern in space when a wavefront moves posteriorly  across the presomitic mesoderm (PSM) from the anterior and freezes the cellular oscillators as it passes by, recording their phases at the time of arrest. Questions of interest  concern the relation between the patterns on the genetic and the cellular levels and the way of how a collective segmentation period rises from the ensemble of individual units. In \cite{juelicher} a delayed coupling theory was developed as a phenomenological  mesoscopic description of the vertebrate segmentation in terms of phase oscillators to represent cyclic gene expression in the cells of the PSM. The oscillators are coupled  to their neighbors with delay and have a moving boundary, describing the elongation of the embryo axis. They are equipped with a profile of natural frequencies that is designed to slow down the oscillations and stop them at the arrest front. In particular it is shown how the clock's collective period depends on the delayed coupling. The very mechanism of arresting the oscillations at the arrest front is, however, not addressed, neither is the onset of oscillations described as a Hopf bifurcation from an underlying dynamical system (as, for example, in \cite{jensen2003} or in \cite{goldbeter2008}). Differently, in \cite{goldbeter2007} and \cite{santillan2008}, the arrest of oscillations is supposed to result from an external signal (rather than from a suitably chosen frequency distribution or a system-inherent bifurcation).

In contrast, in our model of coupled genetic circuits, both the onset and the arrest of oscillations result from two bifurcations as one and the same bifurcation parameter is monotonically increased. Here, beyond a possible later application to the segmentation clock,  we are interested in the more general question: Given a system of coupled units which individually can show excitable or limit-cycle behavior, how can we control the duration  of collective oscillations via bifurcations? The individual units are composed of a positive and a negative feedback loop. A first species $A$ activates itself in a positive feedback loop, and it also activates its own repressor, the second species $B$, in a negative feedback loop. This motif is frequently found in neural and genetic networks in different realizations, whenever bistable units are coupled to negative feedback loops. Examples are signaling systems like
the slime mold Dictyosthelium Discoideum \cite{mold},  the  embryonic division control system \cite{pomerening}, or the MAPK-cascade \cite{mapk}, and the circadian clock \cite{leibler}.

As it was shown in \cite{pablo}, this motif, considered now as an individual unit,  shows three regimes of stationary behavior, when a single bifurcation parameter is monotonically increased: excitable behavior, where the system approaches a fixed point (if the perturbation from the fixed point exceeds a certain threshold, the system makes a long excursion in phase space before it returns to the fixed point), limit-cycle behavior, and again excitable behavior. In \cite{ashok} it was considered in different realizations and termed bistable frustrated unit (BFU). In the following we shall use this abbreviation as well.

When the BFUs are coupled with repressive bonds (being directed, undirected, frustrated or not frustrated), we observe  the same gross collective behavior as the individual units show:  a first regime now of collective fixed-point (CFP)-behavior, followed by collective oscillatory (CO)-behavior, followed by a second regime of CFP-behavior. This sequence is observed for a broad range of parameters, of initial and boundary conditions, when a single, uniformly chosen bifurcation parameter is monotonically increased. Therefore the speed of its variation determines the duration of collective oscillations. The very type of collective transients and stationary states depends on the initial conditions, the boundary conditions, the network topology, and the strength of the repressive couplings, as we shall outline below. How much of the rich transient behavior can actually unfold depends on the speed of variation of the bifurcation parameter.

In view of the segmentation clock our model of coupled BFUs may match this system on an effective mesoscopic level of description, in which the interacting cells are described as coupled BFUs, arising from coupled positive and negative feedback loops on the underlying genetic level. Although first hints point in this direction, it is clearly beyond the scope of this paper to provide an exact mapping between our effective description and the segmentation clock and furthermore to identify the involved proteins and genes in the different realizations of the segmentation clock.

\begin{figure}[ht]
\center
\includegraphics*[width=0.5\columnwidth]{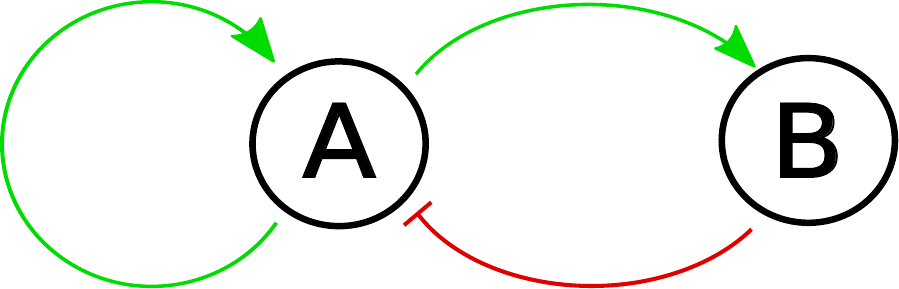}
\caption{Motif of a bistable frustrated unit (BFU). Pointed arrows denote activation (increase in the production rate), the blunt arrow denotes repression (decrease in the production rate).} \label{fig1:motif}
\end{figure}

\section{The model}

\label{sec2}
The motif of a single bistable frustrated unit (BFU) is shown in figure~\ref{fig1:motif}. In this model, $A$ and $B$ are two different species. Although from a statistical physics point of view the identity of $A$, $B$ is not important, we shall assume here that $A$, $B$ are two different protein types. This is in line with the original motivation of this model as a coarse-grained description of some genetic circuits. The protein $A$ activates its own production (transcription in the biological language) and also the production of the protein $B$, which in turn represses the production of $A$. In this way, we have a self-activating bistable unit which is coupled to a negative feedback loop.
The simplest, idealized implementation of the BFU has been analyzed on a deterministic level, with protein concentrations as the only dynamical variables, and it is known to produce oscillations in a certain range of model parameters \cite{guantes,sandeep}.
In particular, the model studied in Ref.~\cite{sandeep}, which we adopt here, assumes that the protein production rates depend on the concentrations $\phi_A$ and $\phi_B$ of the two protein species as follows:
\begin{eqnarray}
 \frac{d\phi_A}{dt}&=&\frac{\alpha}{1+\phi_B/K}\,\frac{b+\phi_A^2}{1+\phi_A^2}-\phi_A, \label{eq-det11} \\
 \frac{d\phi_B}{dt}&=&\gamma(\phi_A-\phi_B), \label{eq-det12}
\end{eqnarray}
where $\gamma$ is the ratio of the half-life of $A$ to that of $B$. Here we shall focus on the case $\gamma\ll 1$, that is when the protein $B$ has a much longer half-life than $A$ with a slow reaction on changes in $A$, while $A$ has a fast response to changes in $B$.  The parameter $K$ sets the strength of repression of $A$ by $B$. We shall assume $K\ll 1$, so that already a small concentration of $B$ will inhibit the production of $A$. The parameter $b$ determines the basal expression level of $A$. We set this parameter to anything larger than zero but much smaller than one, so that the system cannot be absorbed in the state $\phi_A=\phi_B=0$ and, simultaneously, the production rate of $A$ is small for $\phi_A\approx 0$. The Hill coefficients (powers of $\phi_A$, $\phi_B$ on the r.h.s. of Eqs.~(\ref{eq-det11})-(\ref{eq-det12})) are chosen as in \cite{sandeep,pablo}.
The parameter $\alpha$ is the maximal rate of production of $A$ for full activation ($\phi_A^2\gg b$) and no repression ($\phi_B\approx 0$). This will be our tunable parameter which we will use to control the behavior of our model, also when these units are coupled. This parameter seems to be also the easiest one to control in real, experimental systems \cite{sandeep}. The different fixed-point and oscillatory regimes of such an individual unit are separated  by subcritical Hopf bifurcations (for a definition see for example Ref.~\cite{strogatz}) with corresponding hysteresis effects \cite{pablo}.

From now on, we do no longer distinguish between the species and their concentrations in the notation and write $A,B$ for the corresponding concentrations. We consider coupled BFUs with only repressing couplings (for simplicity, since the effect of activating couplings can be compensated by an appropriate choice of geometry, see \cite{pablo}) according to
\begin{eqnarray}\label{eq3}
&&\frac{dA_i}{dt}\;=\;\frac{\alpha}{1+(B_i/K)}\;\cdot\;\left(\frac{b+A_i^2}{1+A_i^2}\right)\;-\;A_i\\ \nonumber
&&+\beta_R\;\sum_{j=1}^{N}R_{ij}\frac{1}{1+(A_j/K)^2}\\
&&\frac{dB_i}{dt}\;=\;\gamma (A_i\;-\;B_i)\;,\qquad i=1,...,N\;.\nonumber
\end{eqnarray}
Here $i$ labels the units. In some figures later we use $A_{i,j}$, in which $i$ and $j$ denote the coordinates in both directions separately to facilitate the localization on the lattice. The parameters $\alpha$, $K$, $b$ and $\gamma$ are chosen uniformly over the grid, and $K$, $b$ and $\gamma$ with the same values as introduced for a single BFU, that is $K=0.02$, $b = \gamma=0.01$, so without any finetuning. The parameter $\alpha$ is also chosen independently of the lattice site, but as a bifurcation parameter it is varied between $1$ and $110$ or $400$, depending on the repressive coupling strength, parameterized by $\beta_R$, being weak or strong, respectively. Throughout the paper we call the coupling $\beta_R=0.1$ weak and $\beta_R=10.0$ strong, $\beta_R$ is chosen with the same value on all bonds and in opposite directions. For the variation of $\alpha$ we distinguish three speeds: fast ($\Delta\alpha=15, \Delta t=100$ time units (t.u.)), gradual ($\Delta\alpha=0.0075, \Delta t=1$ t.u.) and slow ($\Delta\alpha=15 $ or $ 20, \Delta t=4000$ t.u.). The adjacency matrix elements $R_{ij}$ take values 1 if there is a directed coupling from $i$ to $j, i\not=j$ and zero otherwise. The adjacency matrix will be chosen to represent three lattice topologies as explained in the following.

The system (\ref{eq3}) is integrated with the fourth-order Runge-Kutta algorithm with an integration step size of $dt=0.01$, which turned out to be sufficiently small.

\begin{figure}[ht]
\captionsetup[subfigure]{labelformat=simple}
\centering
\subfloat[SQU-OC]
		{\includegraphics*[width=0.18\columnwidth]{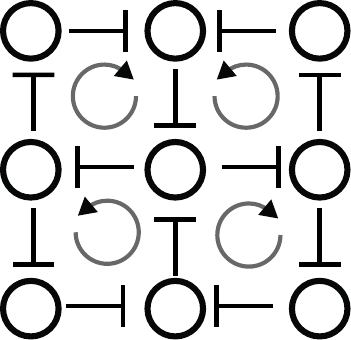}}\hfill
\subfloat[SQU-SC]
		{\includegraphics*[width=0.18\columnwidth]{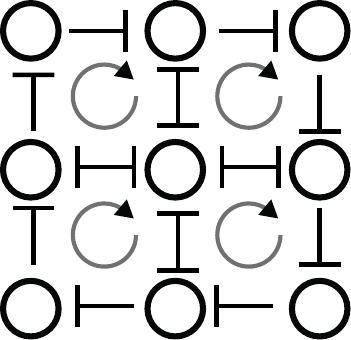}}\hfill
\subfloat[SQU-UD]
		{\includegraphics*[width=0.18\columnwidth]{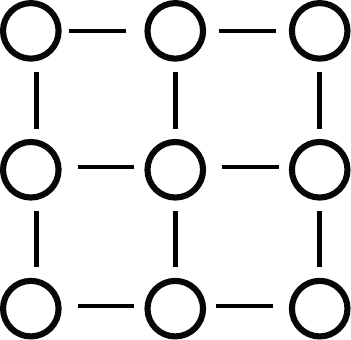}}\hfill
\subfloat[HEX-OC]
		{\includegraphics*[width=0.28\columnwidth]{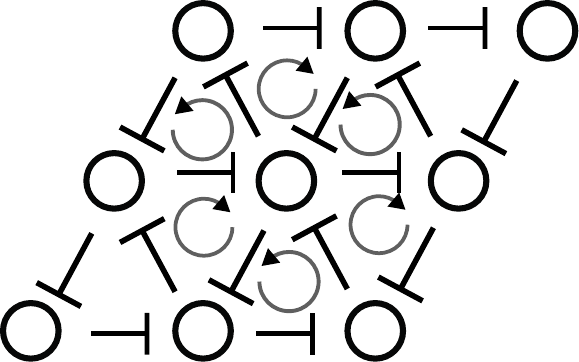}}
\caption{Topologies a) SQU-OC, b) SQU-SC, c) SQU-UD, and d) HEX-OC. At each vertex (circle) we have one BFU repressing two, three, or four neighbors. Blunt arrows represent repressive coupling, while pointed circular arrows represent the flow of repression in the smallest possible loop on the lattice. If two such neighboring loops have a circulation of repression in opposite directions, we say that the lattice has opposite chirality as in a) and d), otherwise the chirality is the same (b)). The lattice c) is undirected and differs from b) only at the boundaries.} \label{fig2:topology}
\end{figure}

\paragraph{Choice of topologies and boundary conditions} In general we choose free (f.b.c.) or periodic (p.b.c.) boundary conditions. For p.b.c. the lattice sites in both directions are identified modulo L, the linear extension.
In order to facilitate the recognition of patterns we choose regular network topologies as indicated in figure~\ref{fig2:topology}: three type of rectangular lattices, usually chosen as squares of size $L\times L$ with $L=50$ apart from some case studies for quasi one-dimensional arrangements of $50\times 2$ up to $1000\times 2$ units, and one hexagonal lattice, again of $L\times L$ units with $L=5$ or $50$.
To have the option of multistable stationary states, we choose lattice and boundary conditions in a way that we have frustrated bonds for specific choices. The criteria for calling bonds frustrated  in directed or undirected excitable media were presented in \cite{pablo}. Accordingly the  geometry of the HEX-OC-lattice of figure~\ref{fig2:topology}d) induces no frustrated bonds, while all bonds are frustrated for the SQU-OC, independently of the choice of boundary conditions, in figure~\ref{fig2:topology}a). An interesting alternative to the square lattice SQU-OC is given by the SQU-SC and SQU-UD-lattices, see figure~\ref{fig2:topology} b) and c). Lattice SQU-UD has no frustrated bonds for p.b.c. and all couplings chosen with the same strength in both directions of traversing a bond (so the bonds become effectively undirected), see figure~\ref{fig2:topology} c). For f.b.c., we choose the bonds along the boundary to be directed, but undirected in the bulk, as displayed in figure~\ref{fig2:topology} b). Apart from inducing frustration to certain bonds, the lattice SQU-SC then has an overall non-vanishing chirality, and all elementary plaquettes have the same chirality when each bond in the bulk is traversed twice, but in opposite directions, as indicated in figure~\ref{fig2:topology} b). In contrast, the SQU-OC and the HEX-OC have adjacent plaquettes with opposite chirality. These apparently minor features seem to be relevant for the patterns and their sensitivity to the boundary conditions.

In view of biological applications it is questionable whether multistability is a desired feature. Here we choose the different topologies with and without frustrated bonds and different chirality in order to illustrate the common features and the differences in the transient and stationary collective behavior of the whole system. The very choice of regular topologies may be even not unrealistic in view of cell assemblies in biological systems \cite{jensen2009}.

\begin{figure}[ht]
\captionsetup[subfigure]{labelformat=simple}
\centering
	\subfloat[]
		{\includegraphics*[width=.230\columnwidth]{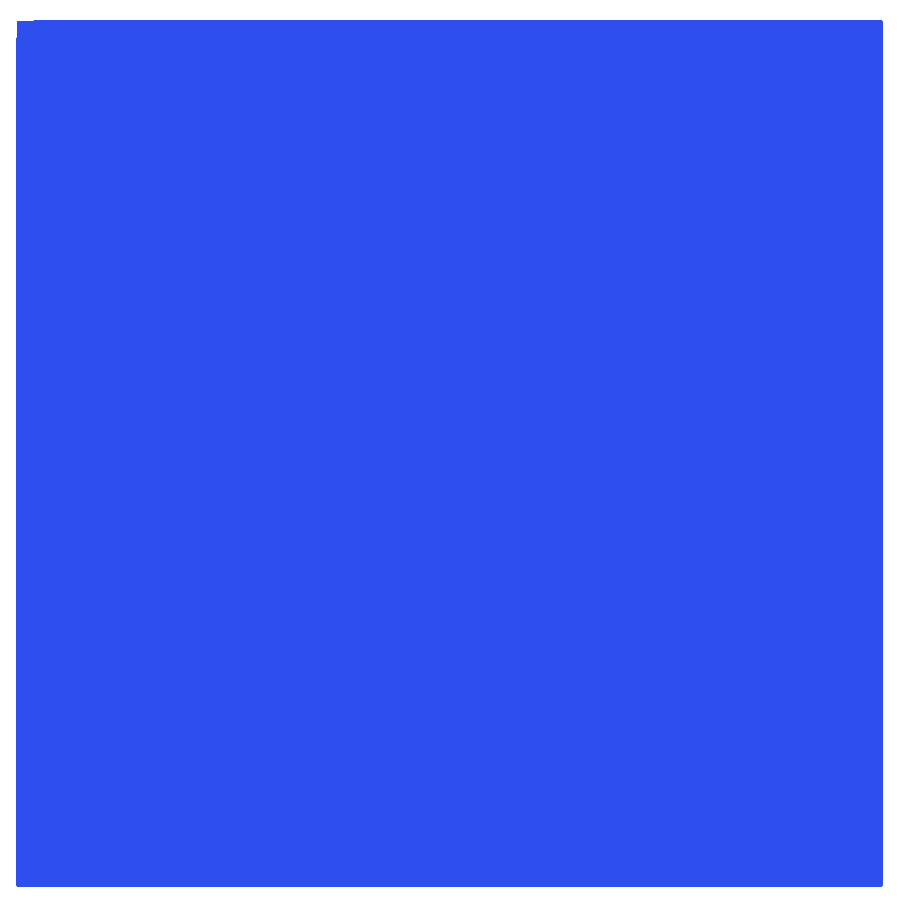}}\hfill
	\subfloat[]
		{\includegraphics*[width=.230\columnwidth]{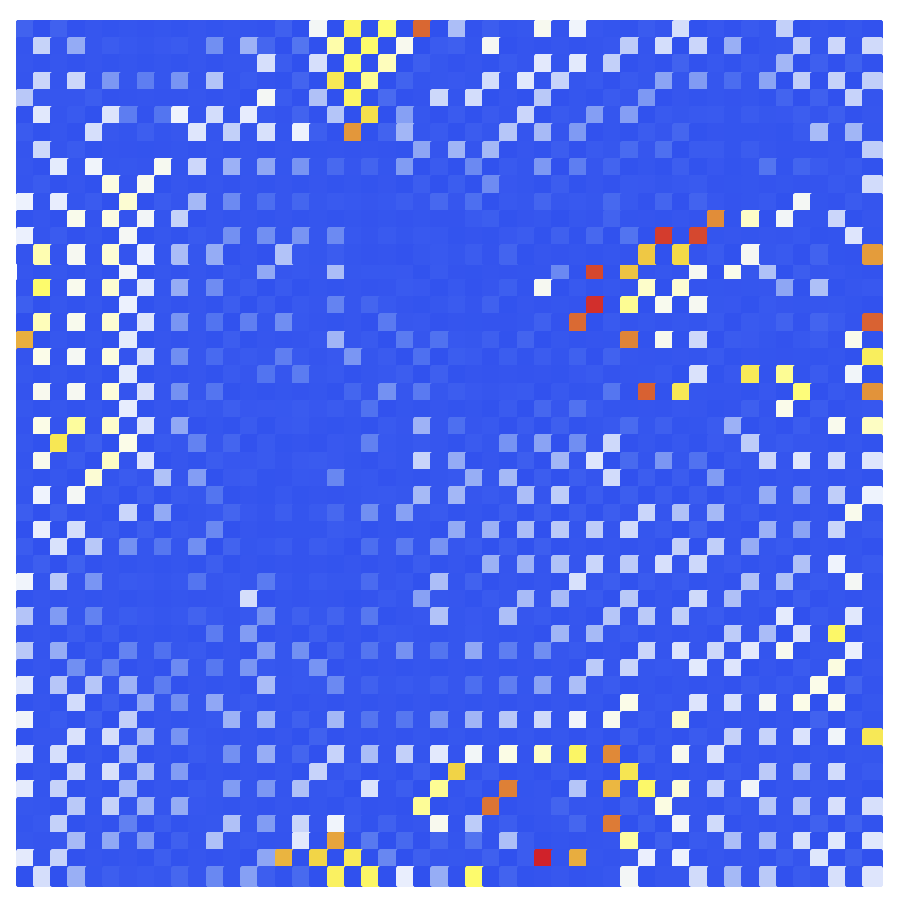}}\hfill
	\subfloat[]
		{\includegraphics*[width=.230\columnwidth]{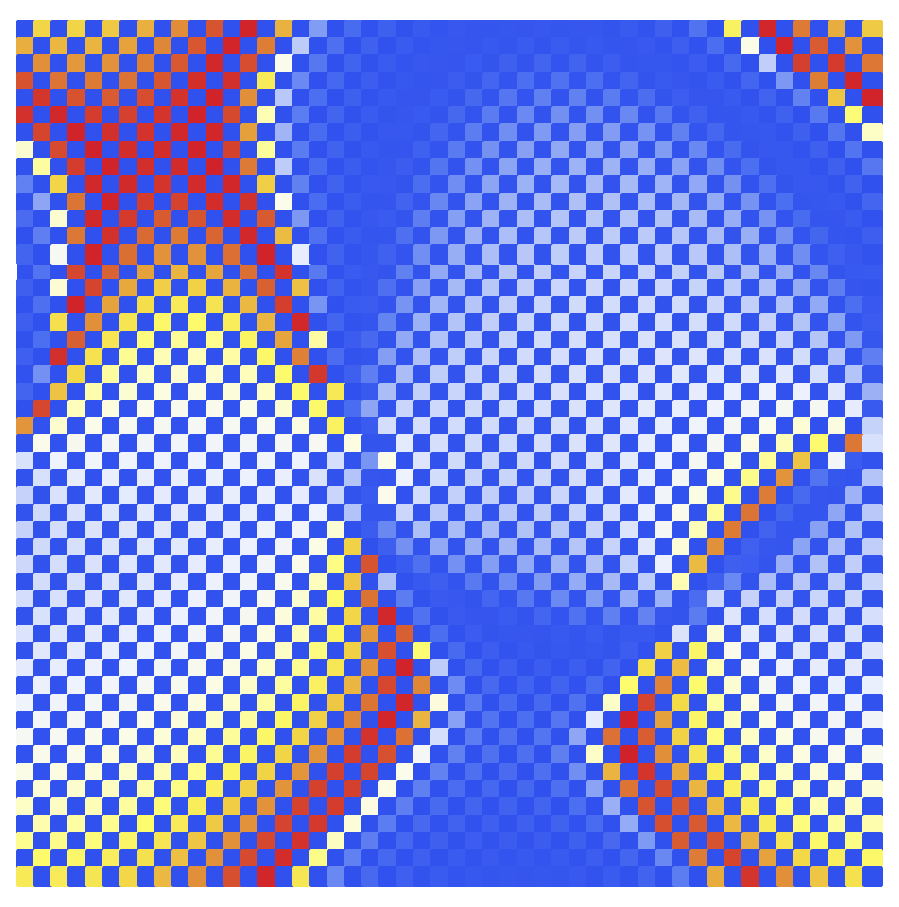}}\hfill
	\subfloat[]
		{\includegraphics*[width=.230\columnwidth]{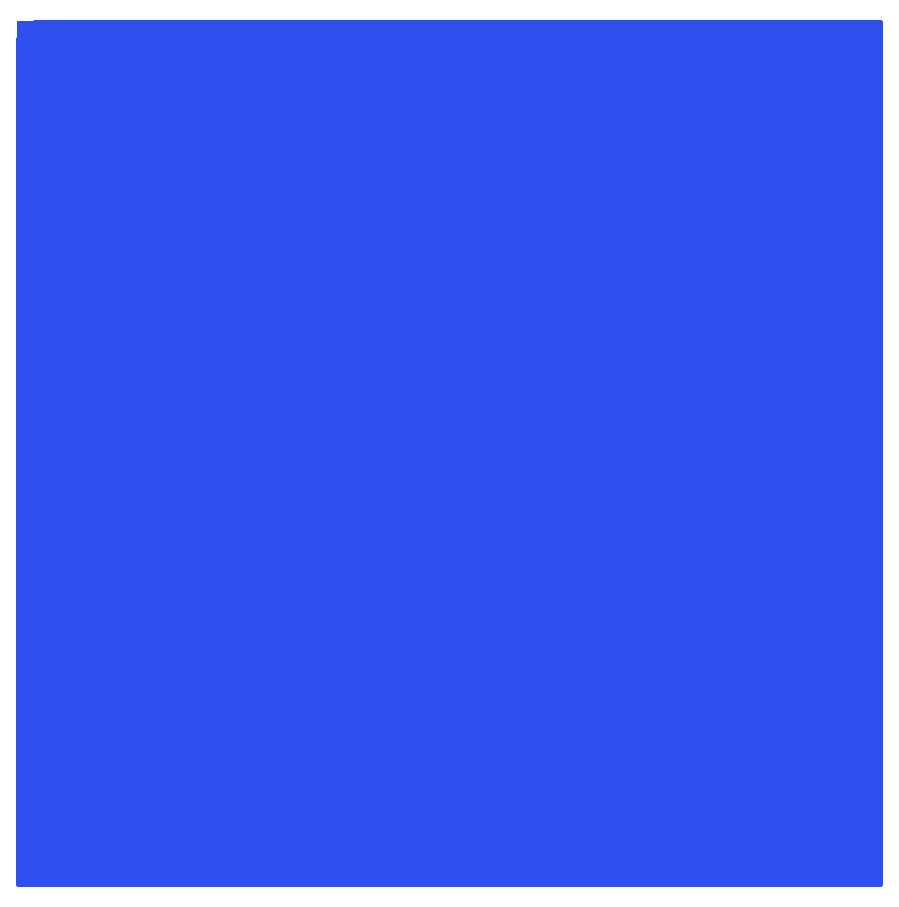}}\\
	\subfloat[]
		{\includegraphics*[width=.230\columnwidth]{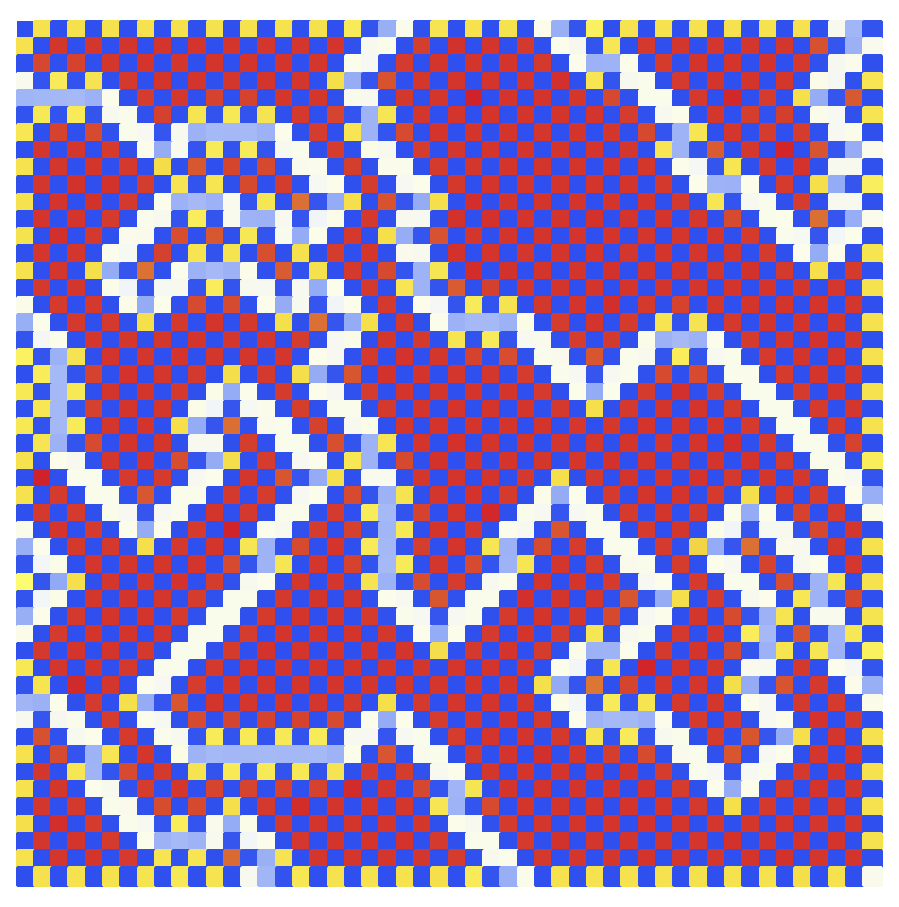}}\hfill
	\subfloat[]
		{\includegraphics*[width=.230\columnwidth]{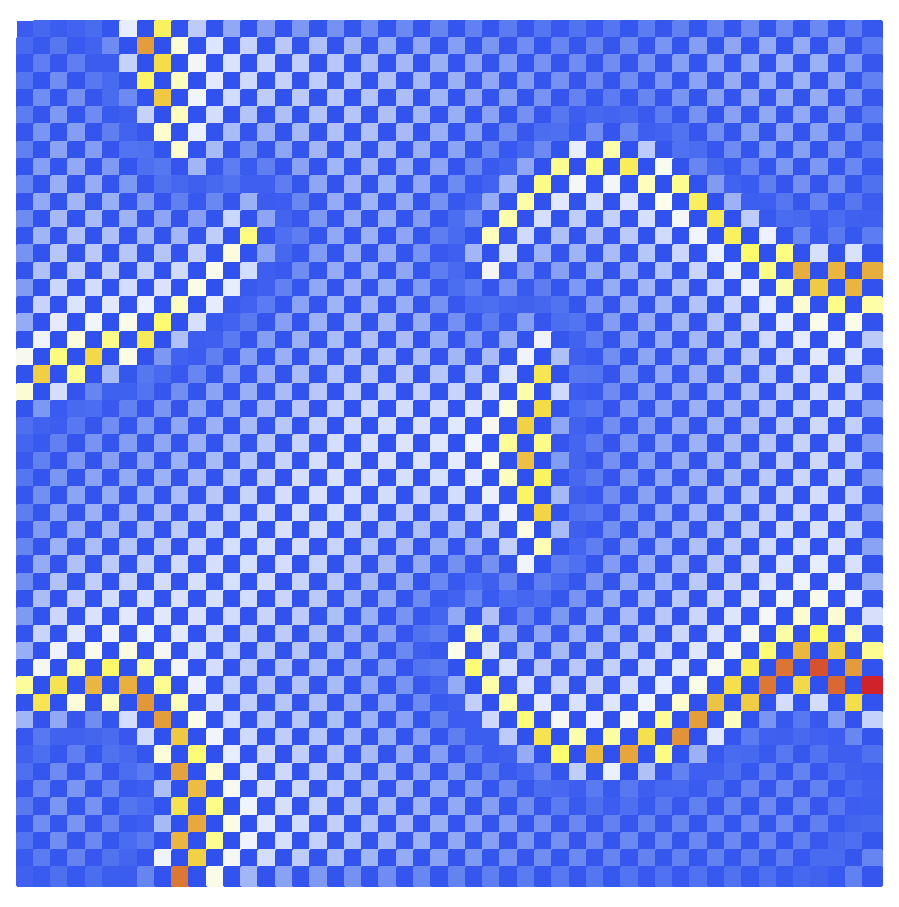}}\hfill
	\subfloat[]
		{\includegraphics*[width=.230\columnwidth]{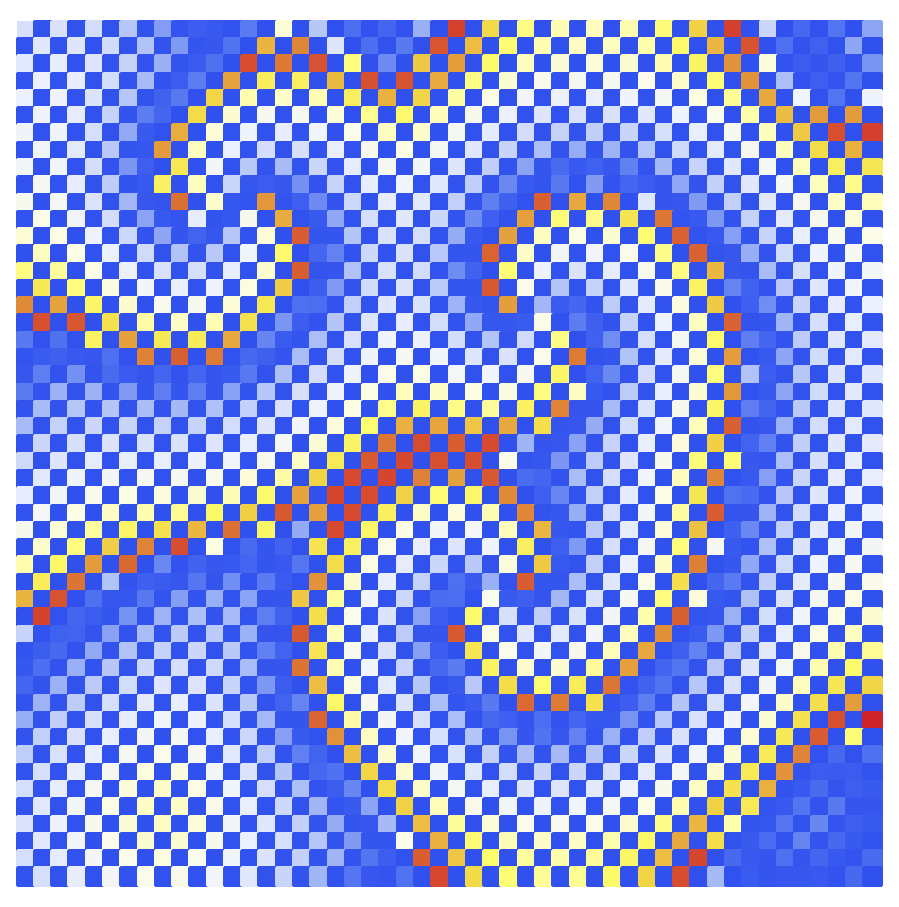}}\hfill
	\subfloat[]
		{\includegraphics*[width=.230\columnwidth]{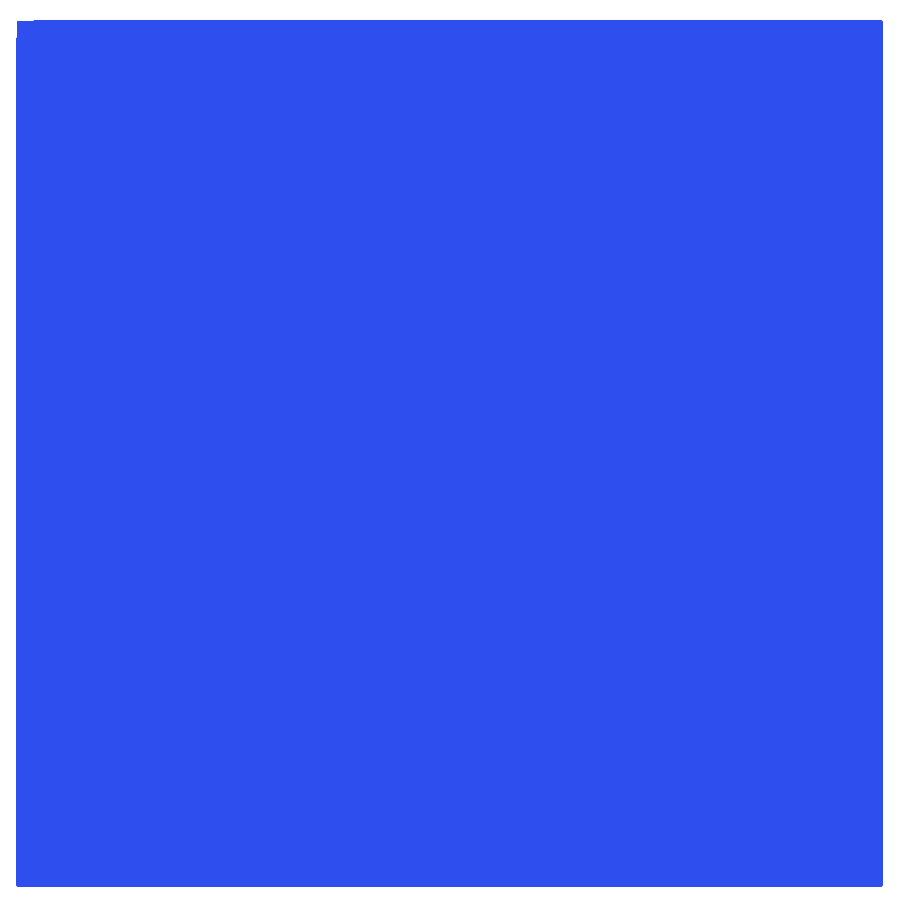}}\\
	\subfloat[]
		{\includegraphics*[width=.230\columnwidth]{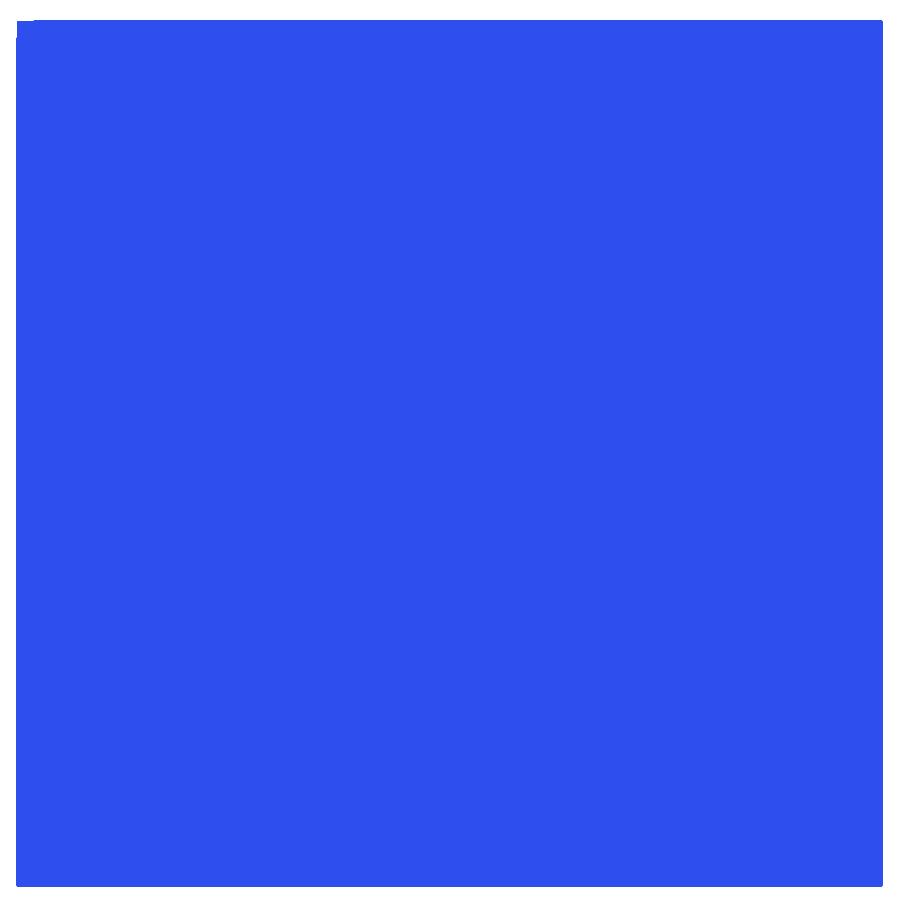}}\hfill
	\subfloat[]
		{\includegraphics*[width=.230\columnwidth]{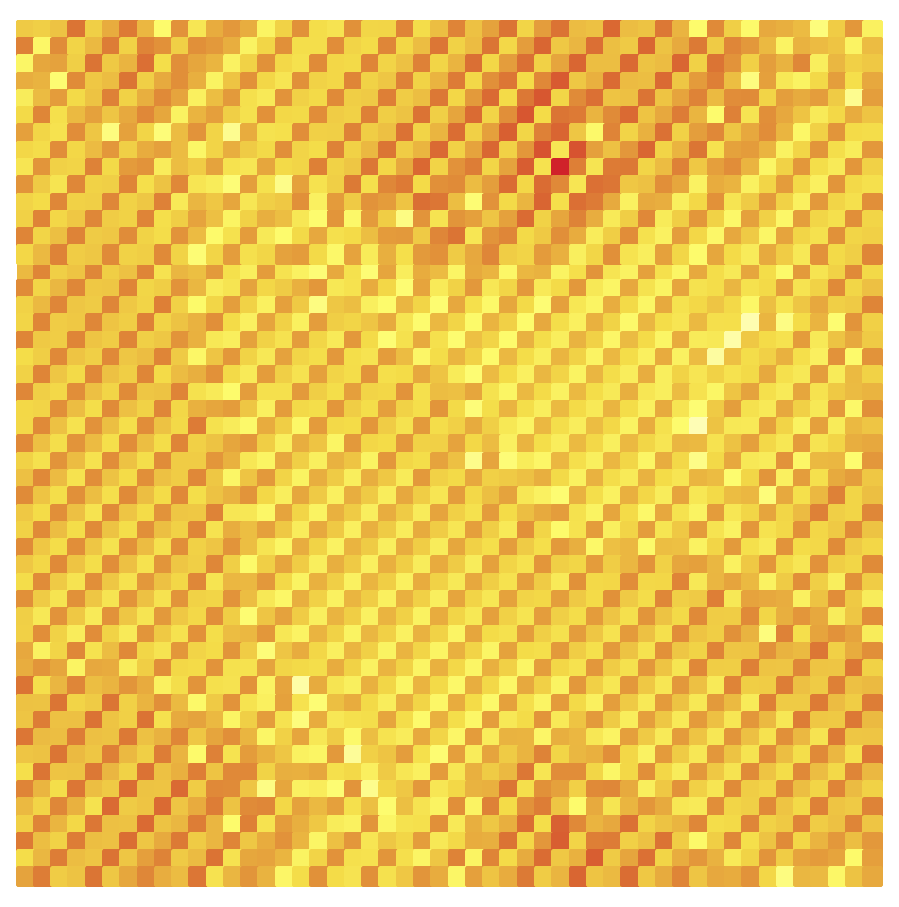}}\hfill
	\subfloat[]
		{\includegraphics*[width=.230\columnwidth]{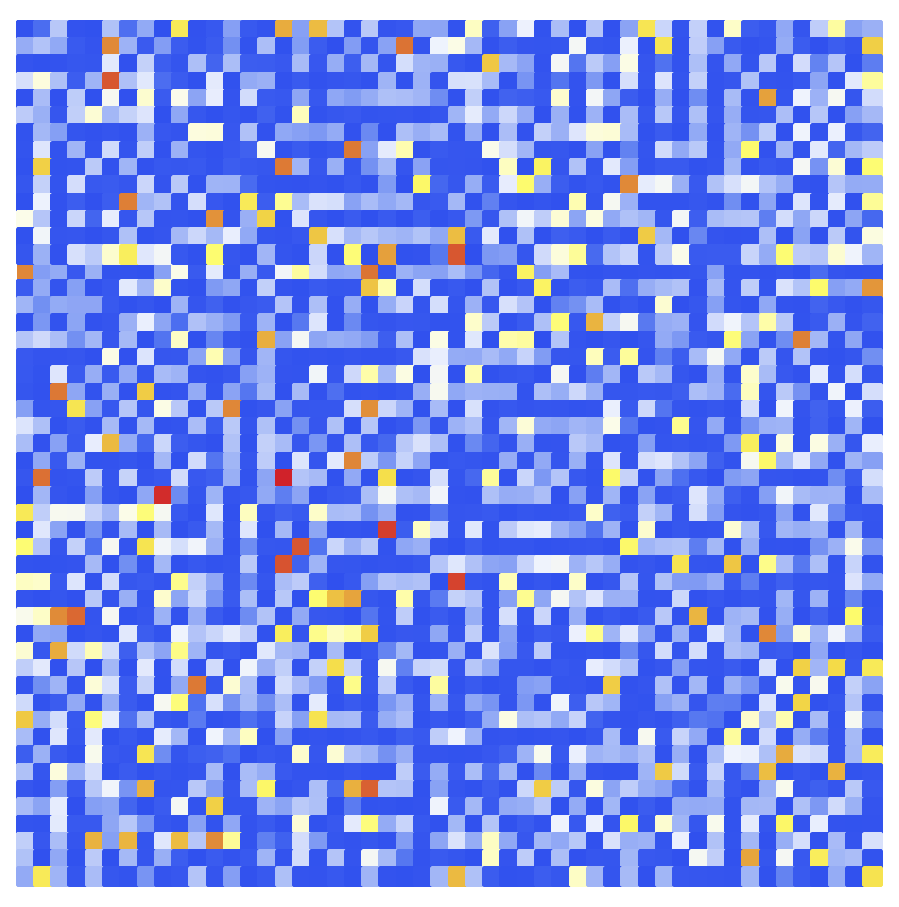}}\hfill
	\subfloat[]
		{\includegraphics*[width=.230\columnwidth]{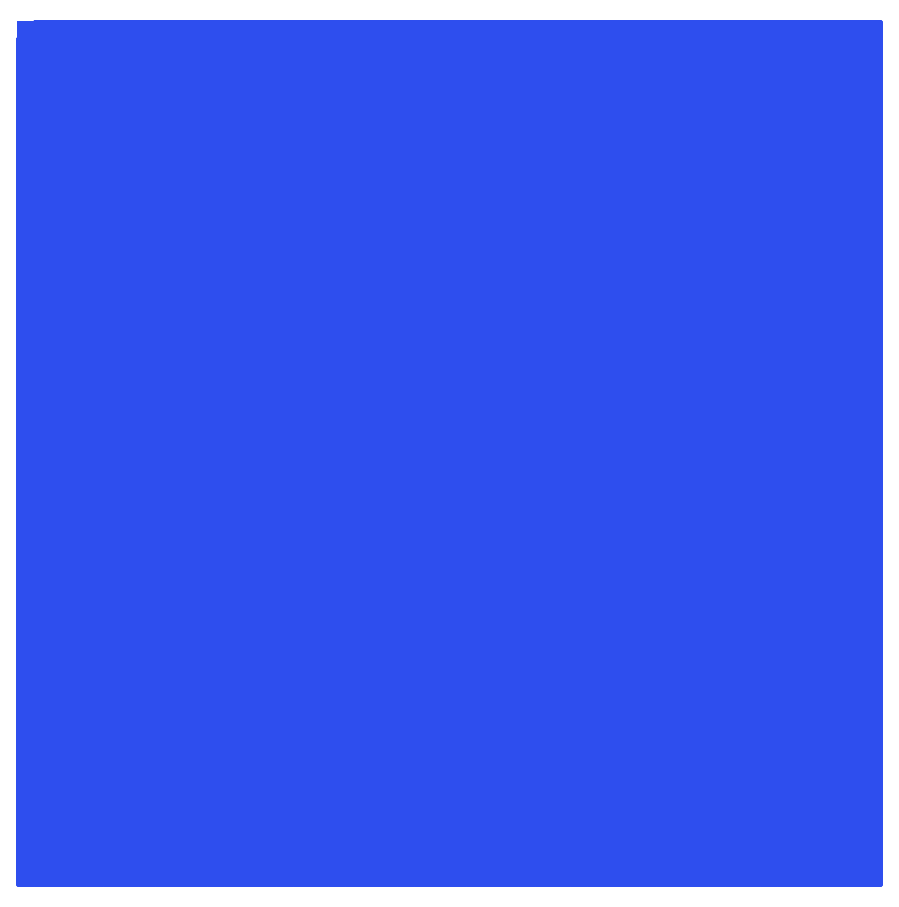}}
\caption{Snapshots of collective fixed-point behavior (left column), two kinds of collective oscillatory behavior (second and third column), and again collective fixed-point behavior (fourth column) for three different topologies: SQU-OC with p.b.c. (first row), SQU-UD with f.b.c. (second row), and HEX-OC with p.b.c.(third row). All simulations are done for gradual variation of $\alpha$, increasing from left to right in the panels, and weak coupling. For further explanations see the text.} \label{fig3:fp-osc-fp}
\end{figure}

\section{Results}

\subsection{Control of the duration of oscillations}
Before we go into more detail on the dependence of the lattice topology, the coupling strength and the variation speed of the bifurcation parameter $\alpha$, we summarize the features which are in common to almost all realizations of our system. While we know from the dynamics of an individual unit that it shows three regimes when the bifurcation parameter $\alpha$ is monotonically  varied, it may come as a surprise that similar kind of regimes  emerge as collective behavior, when the individual units are coupled and the very same control parameter is monotonically tuned. Note that it is again a single, now uniformly chosen control parameter $\alpha_i=\alpha$ for all $i=1,...,N$ that controls the duration of the intermediate regime of collective oscillations (CO). In the stationary state, the collective oscillations are synchronized in frequency, but differ by their patterns of synchronized phases. Now, for small and large values of $\alpha$ we have collective fixed-point behavior (CFP), in which all individuals approach a fixed point in concentration. This sequence of CFP-CO-CFP is seen for all lattice topologies, values of variation speed and couplings, see figure~\ref{fig3:fp-osc-fp}, apart from a few exceptions.
Figure~\ref{fig3:fp-osc-fp} shows snapshots of the values of $A$ for three topologies: SQU-OC with p.b.c. (first row), SQU-UD with f.b.c. (second row), and HEX-OC with p.b.c. (third row) at time instants from the CFP-regime (first and fourth column) and out of the CO-regime (second and third column). All simulations were run for a gradual variation of $\alpha$  and at weak coupling, while the other parameters were kept fixed ($k=0.02$, $b=\gamma=0.01$). Here and throughout the paper, the color codes the value of $A$: ranging from blue for zero, to red for a maximum value.

The collective fixed point and oscillatory behavior can be further characterized by the number of clusters, where we collect in one cluster  all units $i$ for which $A_i$ ($B_i$) agree within the numerical accuracy. In this sense the uniformly blue panels represent a one-cluster fixed point, panel e) a four-cluster fixed point, and the second and third columns indicate multi-cluster limit cycles with more or less regular patterns. While the displayed fixed-point behavior  corresponds to the stationary behavior at these $\alpha$-values, the columns of the oscillatory regimes represent transient patterns upon approaching the stationary state. The snapshots were taken  after 2500, 6000, 10000 and 15000 t.u. (first row), 100, 1500, 10000, and 15000 t.u. (second row), and 3000, 4000, 12000, and 15000 t.u. (third row). A typical time evolution of a set of $50\times 50$ units on a SQU-UD lattice for a gradual variation of $\alpha$ and weak coupling is shown in the movie \cite{mov:squ_ud_weak}\footnote{Captions to the movies can be found in the following link \href{https://dl.dropboxusercontent.com/u/97442991/captions_to_the_movies.pdf}{captions.pdf} }.

The first exception is a HEX-OC for strong coupling: instead of a CFP and CO-regime we find interchanging intervals of chaotic and periodic oscillations for small and intermediate  values of $\alpha$ (where the patterns of the periodic intervals are traveling waves), followed by a CFP-regime for large $\alpha$ as displayed in the movie \cite{mov:hex_strong}. The second exception is a SQU-OC-lattice for weak coupling $\beta_R$, where we find two additional regimes for small $\alpha$, such that the total sequence is given by CFP-CO-CFP-CO-CFP \cite{mov:squ_oc_weak}, characterized by a two-cluster CFP, two-cluster CO, one-cluster CFP, multiple-cluster CO, and one-cluster CFP, respectively. The second and third additional regimes disappear for strong coupling. %The first two additional regimes disappear for strong coupling.

\begin{figure}[ht]
\captionsetup[subfigure]{labelformat=simple}
\centering
\subfloat[]
		{\includegraphics*[width=.230\columnwidth]{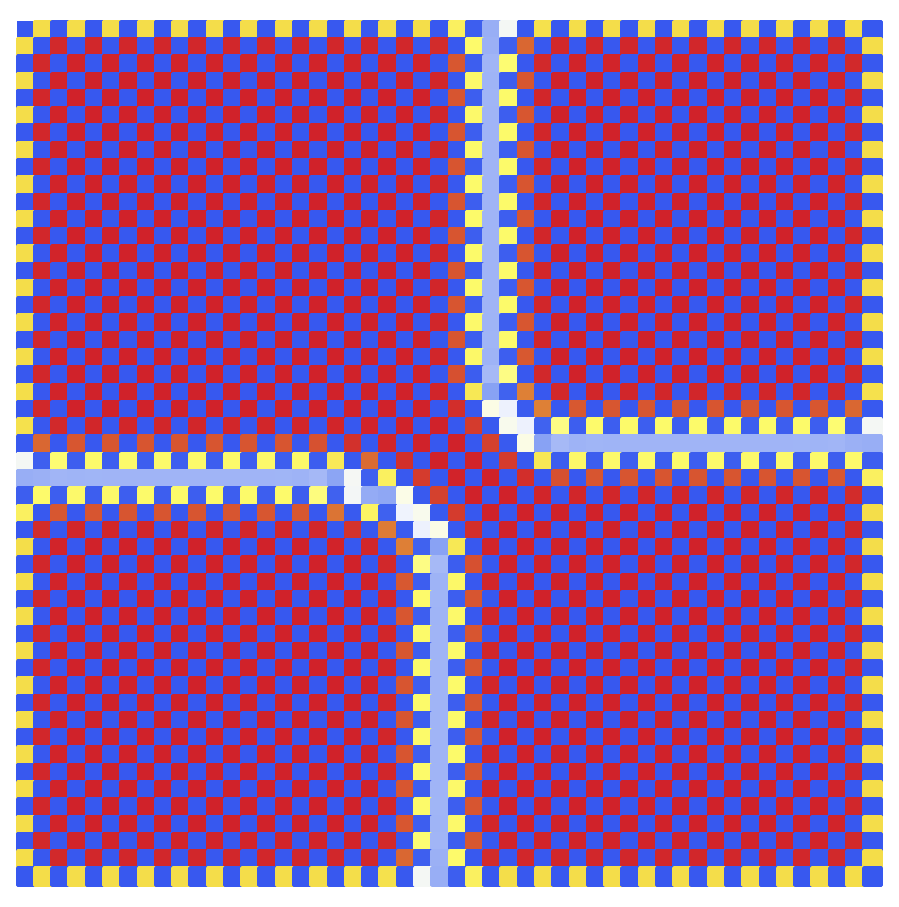}}\hfill
\subfloat[]
		{\includegraphics*[width=.230\columnwidth]{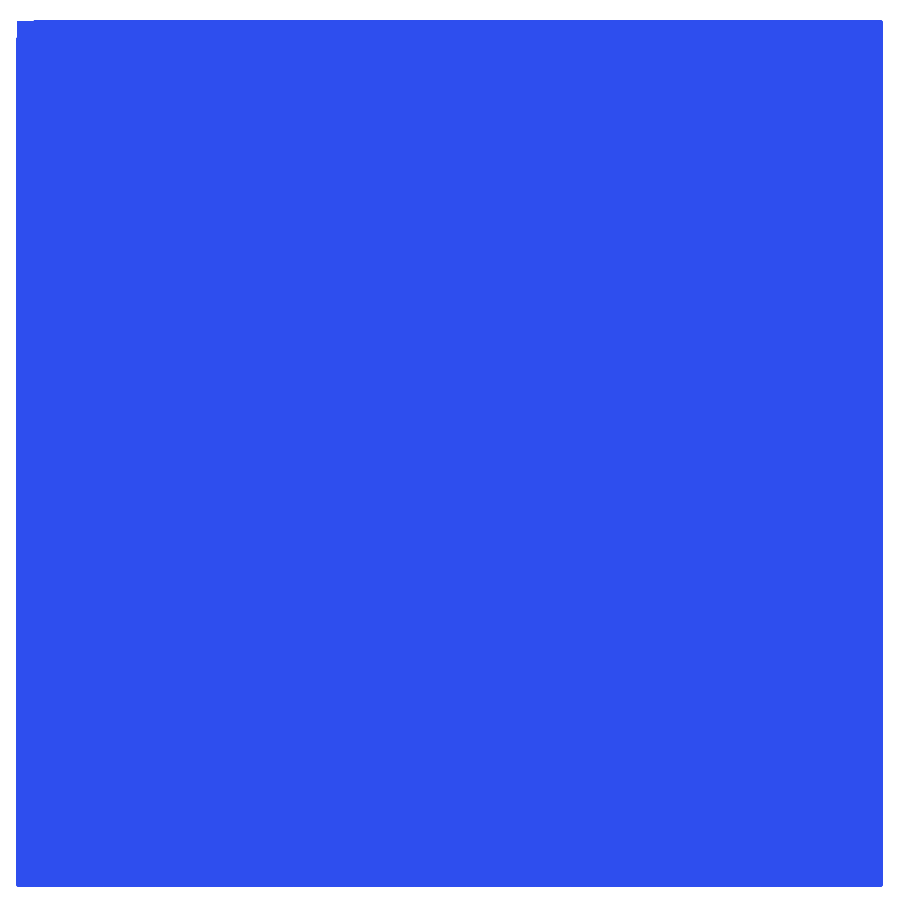}}\hfill
\subfloat[]
		{\includegraphics*[width=.230\columnwidth]{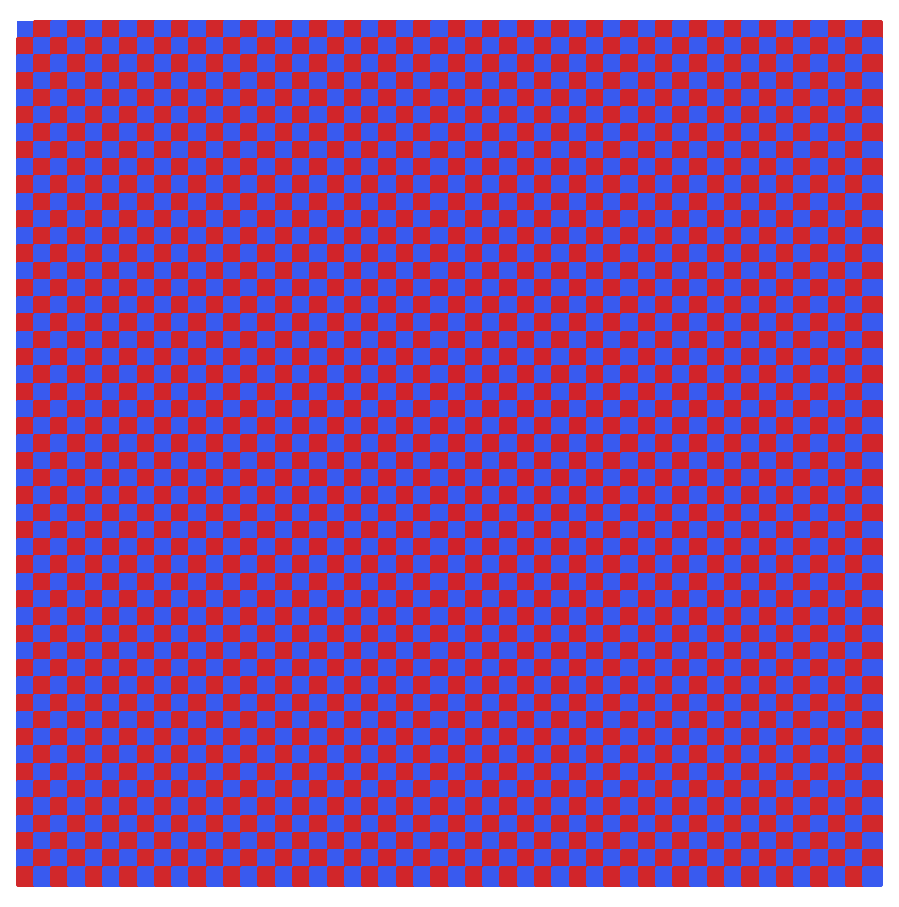}}\hfill
\subfloat[]
		{\includegraphics*[width=.230\columnwidth]{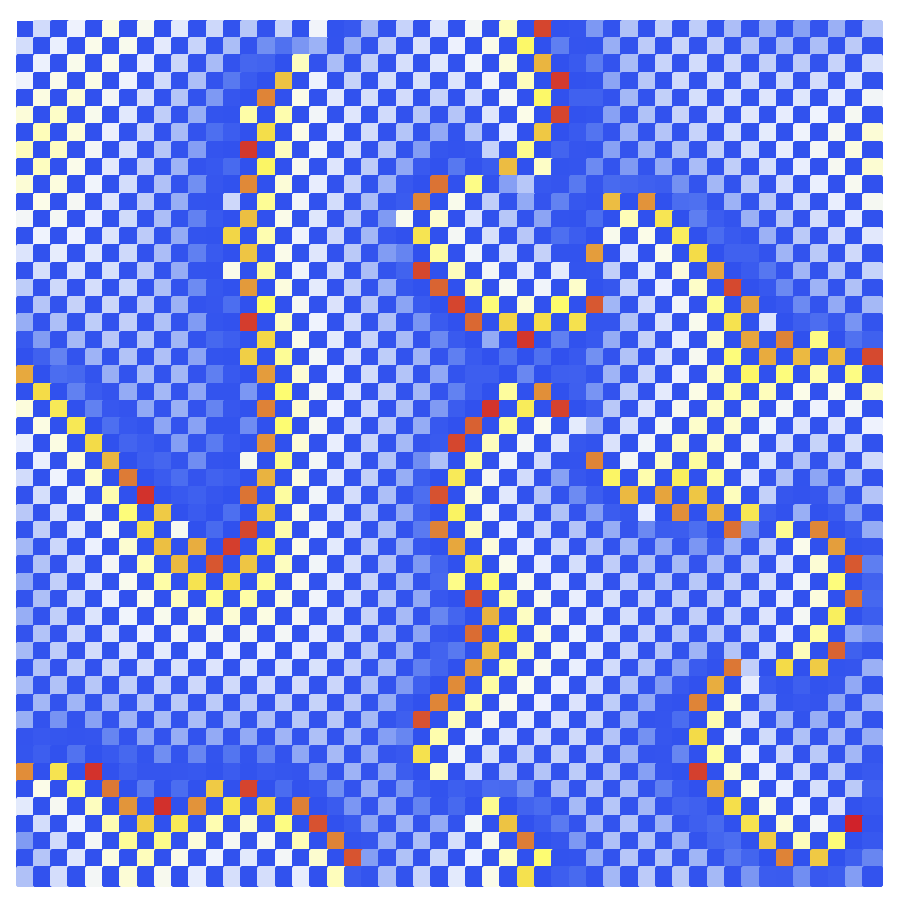}}
\caption{Examples for stationary states for fixed $\alpha$ and weak coupling. From left to right: multi-cluster fixed point on a SQU-UD at $\alpha=1.0$ after 5000 t.u., starting from ``ordered" initial conditions $A_i = B_i = 0.5$, and $A_3 = 0.0$; one-cluster fixed point (SQU-OC), $\alpha=110.0$ and p.b.c. after 5000 t.u. with random initial conditions; two-cluster limit cycle and multi-cluster limit cycle on a SQU-OC lattice, for $\alpha = 90.0$ and ordered (c))versus random (d)) initial conditions. }\label{fig4:stat}
\end{figure}

For demonstrating the main control mechanism there was no need to await the approach of stationary states, before $\alpha$ was changed. Therefore, we briefly summarize the different type of stationary states without entering details about the choice of parameters. In the CPF-regimes we have time-independent states, in which the concentrations approach fixed-point values as one-cluster (figure~\ref{fig4:stat} b)), two-cluster (not displayed), or multi-cluster fixed points (figure~\ref{fig4:stat} a)), depending on the lattice topology and the initial conditions. In the CO-regime, the solutions are frequency-synchronized oscillations of individual units, arranged in two-(figure~\ref{fig4:stat} c))or multi-cluster distributions (figure~\ref{fig4:stat} d)). In the latter case, the spiral patterns created in the transient phase, are preserved. For a HEX-OC lattice and strong coupling we see chaotic time evolutions
of concentrations rather than the usual sequence of regimes, cf. the discussion in section~\ref{secstrong}.

In general, for the $50\times 50$ lattices and fixed $\alpha$, the transient time to approach a fixed point is of the order of some hundred to thousand time units, while it can take up to 200000 time units to approach the synchronized limit cycles.

\subsection{Transient patterns such as self-organized pacemakers}\label{secpace}
The type and duration of  transient patterns depends on
\begin{itemize}
\item
the topology. We see spirals for all square lattices and no recognizable regular structure for the HEX-OC, see below.
\item
the boundary conditions. Rectangular waves are observed for f.b.c. and stripes for p.b.c. on the SQU-SC-lattice.
\item
the initial conditions. For all square lattices we observed multistability either only in an oscillatory regime (SQU-OC) or both in a first CFP- and CO-regime (SQU-SC and SQU-UD). In these regimes it therefore depends on the initial conditions to which attractor the system will evolve, and consequently which patterns will be generated. If we choose for a hexagonal lattice (HEX-OC) all phases to be initially the same and start with $\alpha$ in the oscillatory regime, the system oscillates with fully synchronized frequencies and identical phases. This limit cycle is, however, unstable: If we slightly perturb the state by kicking one phase by a difference of $0.0004$, the former collective uniform limit cycle is lost and the system evolves to a multi-cluster limit cycle. If we then start with $\alpha$ in the CFP-regime, where the fixed points are reached within only $\sim 1000$ time units, and vary $\alpha$ slowly or gradually, the system has reached the fixed point upon approaching the CO-regime. With these initial conditions the phases remain the same for the rest of the simulation. This is the reason why we do not see pattern formation for a HEX-OC-lattice for a slow change of $\alpha$. In contrast, for fast and gradual change of $\alpha$ and randomly chosen initial conditions, we do see irregular patterns in the CO-regime.
\end{itemize}
The large variety of transient patterns that is seen upon varying $\alpha$ is due to the qualitative diversity of the initial conditions: while we have mainly chosen two type of initial conditions for fixed $\alpha$, random and or ordered ones, i.e. identical ones, perturbed at a single site, the initial conditions for varying $\alpha$ are in general neither random nor highly-ordered, but structured by the pattern that was reached before the $\alpha$-value was increased.  So it is the variety of transient patterns, which set the subsequent initial conditions, and the sensitivity to the initial conditions is in general pronounced, as we have multistable states as attractors with sufficiently large basins of attraction.

Apart from spirals, rectangular or plane waves, we see also target waves, emitted from units that act like dynamically generated pacemakers, this means in a self-organized way. The phenomenon is seen for a small range of speeds of increasing $\alpha$ for the SQU-OC and p.b.c., and for some values of fixed $\alpha$ as a transient towards a two-cluster limit cycle. So it is less generic than the generation of spirals, but remarkable as a transient.

\begin{figure}[ht]
\captionsetup[subfigure]{labelformat=simple}
\centering
	\subfloat[]
		{\includegraphics*[width=.230\columnwidth]{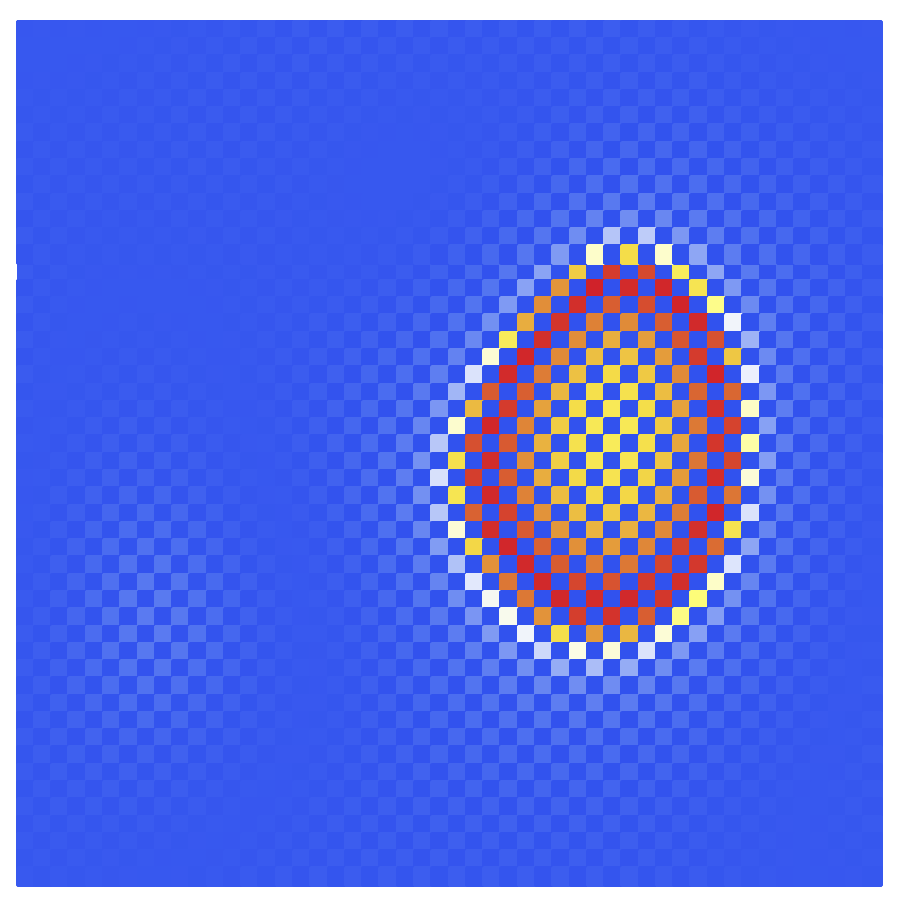}}\hfill
	\subfloat[]
		{\includegraphics*[width=.230\columnwidth]{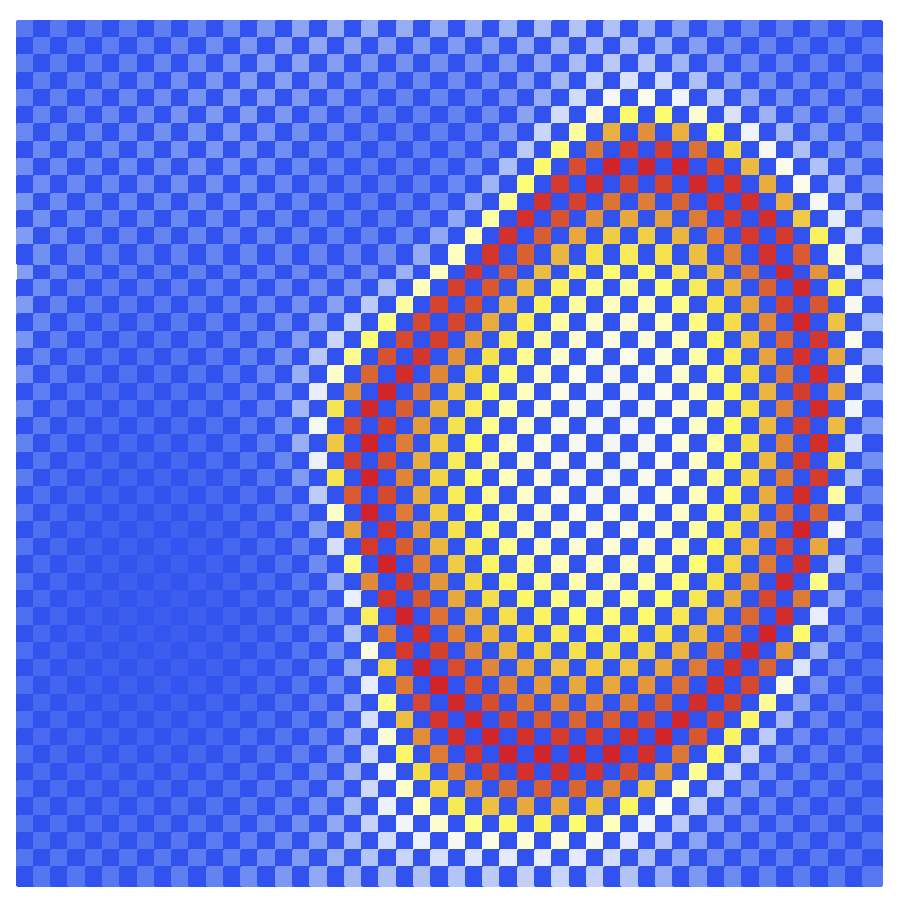}}\hfill
	\subfloat[]
		{\includegraphics*[width=.230\columnwidth]{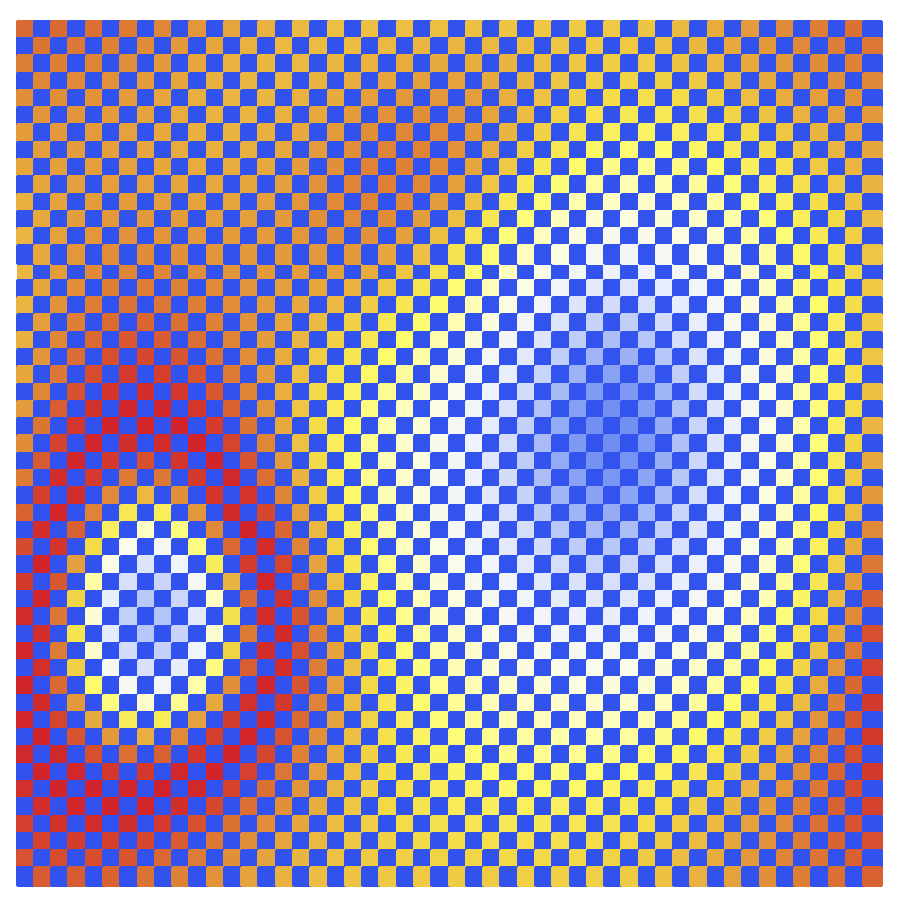}}\hfill
	\subfloat[]
		{\includegraphics*[width=.230\columnwidth]{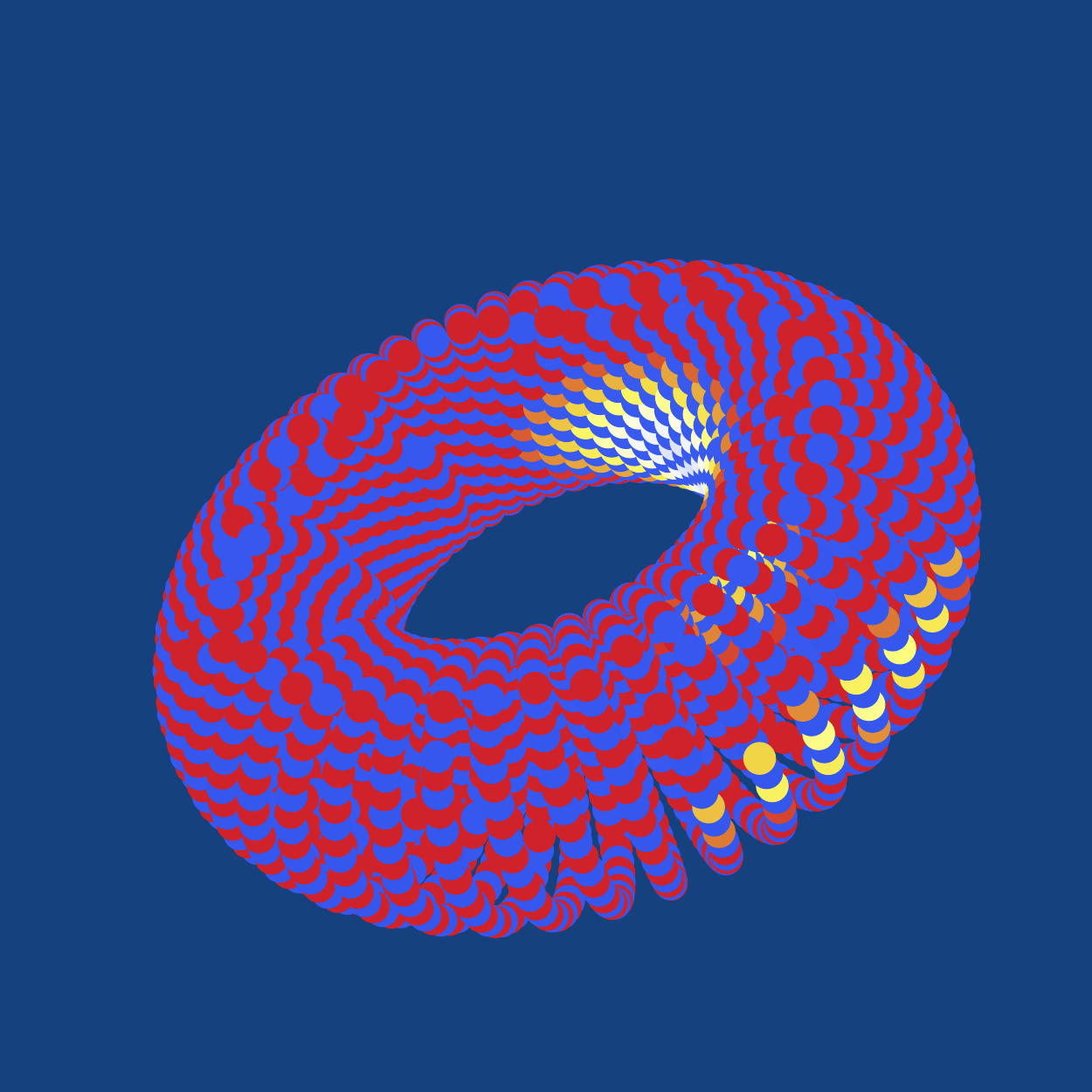}}
\caption{Snapshots of a pacemaker obtained on a SQU-OC lattice with p.b.c. for slow variation of $\alpha$ and weak coupling at times 14010 (a), 14020 (b), and 14030 (c) and (d), when the value of $\alpha$ was 50. Panel (a) shows a source of a circular wave that grows (b) and later shrinks into a sink. Both a sink and a source can be seen in (c) and in the corresponding 3d-plot of (d). } \label{fig5:pacem}
\end{figure}

In an ensemble of oscillators, pacemakers are those units, which entrain the phases of other oscillators to a synchronized oscillation  in space and time such that the pacemakers become the centers of concentric target waves.  Pacemakers here are observed for  the SQU-OC-lattice with p.b.c. as transient either when $\alpha$ is fixed ($\alpha=6$) and $\beta_R$ is weak, in the first CO-regime of the SQU-OC, or when $\alpha$ is slowly varied  at weak coupling. This choice of speed within a small interval of allowed variations is essential for the observation.
On the torus topology we then have two centers opposite to each other, one acting as a source of waves, the other one as their sink. The snapshots of figure~\ref{fig5:pacem} display the phase differences of oscillators with respect to the maximal phase value at the time instant of the snapshot. We see roughly circular spots of the same phase differences, located in the bulk of identical phases (blue background in figure~\ref{fig5:pacem} a)-b)). The shrinking of the spot corresponds to a sink with respect to the front of constant phase differences, its extension to a source of the same front. In the snapshot of figure~\ref{fig5:pacem} c)  we see a time instant, for which the two spots, corresponding to the source and the sink with maximal and minimal phase differences are visible at the same time, the corresponding three-dimensional plot is shown in ~figure~\ref{fig5:pacem} d). The movie \cite{mov:pacem} shows an evolution of few periods on a torus. Altogether this collective arrangement of phases lasts over several thousand time units, before this transient approaches a two-cluster CO-state in the form of a chessboard pattern as the stationary state.

Rather remarkable about this phenomenon is its emergence in a set of units with a completely uniform  choice of individual parameters ($b,K,\gamma,\alpha$) and coupling $\beta_R$. Usually pacemakers are implemented with an ad hoc distinction  via their natural frequencies (see, for example \cite{filippo1}) or some gradient in the natural frequencies of all oscillators \cite{filippo2}; in those cases the location of the pacemaker is predetermined by an ad-hoc implemented distinguished natural frequency. In contrast, our natural frequencies are induced  by a completely uniform choice of parameters in the underlying dynamics, neither is the final location of the emerging spot distinguished in view of the p.b.c., but the very occurrence of  a pacemaker breaks the symmetry between the oscillatory units, so that their heterogeneity must be dynamically generated. Our system, in spite of all couplings being repressive, is of the activator-inhibitor type, as the repression of a repressing bond acts effectively activating. Similar self-organized pacemakers have been identified in reaction-diffusion systems \cite{stich1,stich2,stich3}, there, however, as stable phenomena. In those systems the occurrence of self-organized pacemakers could be traced back to the vicinity of a supercritical pitchfork-Hopf bifurcation, in which a difference in frequencies is dynamically generated in this bifurcation. In our case, the conditions for the observation of these transient phenomena and the origin of their instability in terms of bifurcations should be further analyzed in future work.
Although our self-organized pacemakers are transients, their duration over several thousand time units may be sufficient for providing a signal in the biological context before the system stabilizes to a stationary state.

\subsubsection{A zoom into the transition regime}
For the stationary states we observed two qualitatively different regimes with CFP and CO-behavior. It is of particular interest, both from the viewpoint of statistical physics and of nonlinear dynamics, how the transition between the different collective behavior happens in detail. A detailed bifurcation analysis for a system of the size of $50\times 50$ units would be rather involved (so far we performed a detailed analysis for a system of only two mutually repressing units \cite{forthcoming}). Here we want to adopt the perspective from statistical physics and analyze the emergence of individual fixed-point behavior in the bulk of oscillatory behavior and vice versa, depending on the direction of change of the bifurcation parameter. So the first question is about a possible transient coexistence of both kind of behavior, similar to the coexistence of liquid and gaseous phases of a substrate for a given temperature and pressure.

\begin{figure}[ht]
\captionsetup[subfigure]{labelformat=simple}
\centering
	\subfloat[]
		{\includegraphics*[width=0.23\columnwidth]{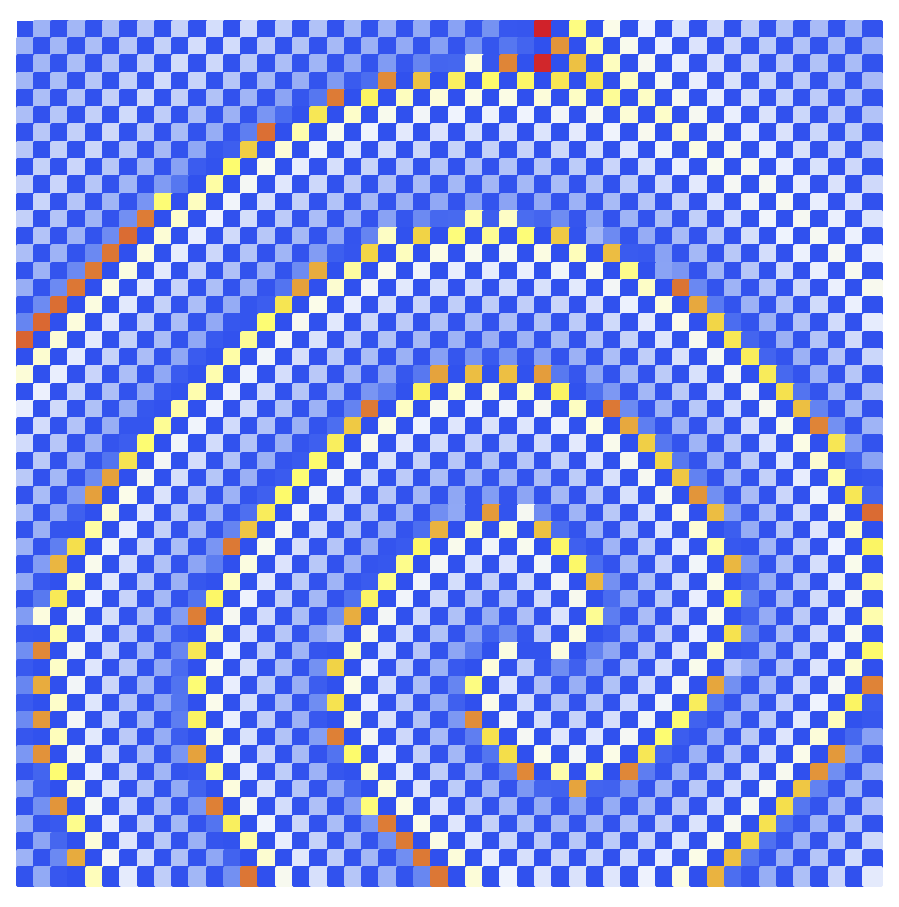}}\hfill
	\subfloat[]
		{\includegraphics*[width=0.23\columnwidth]{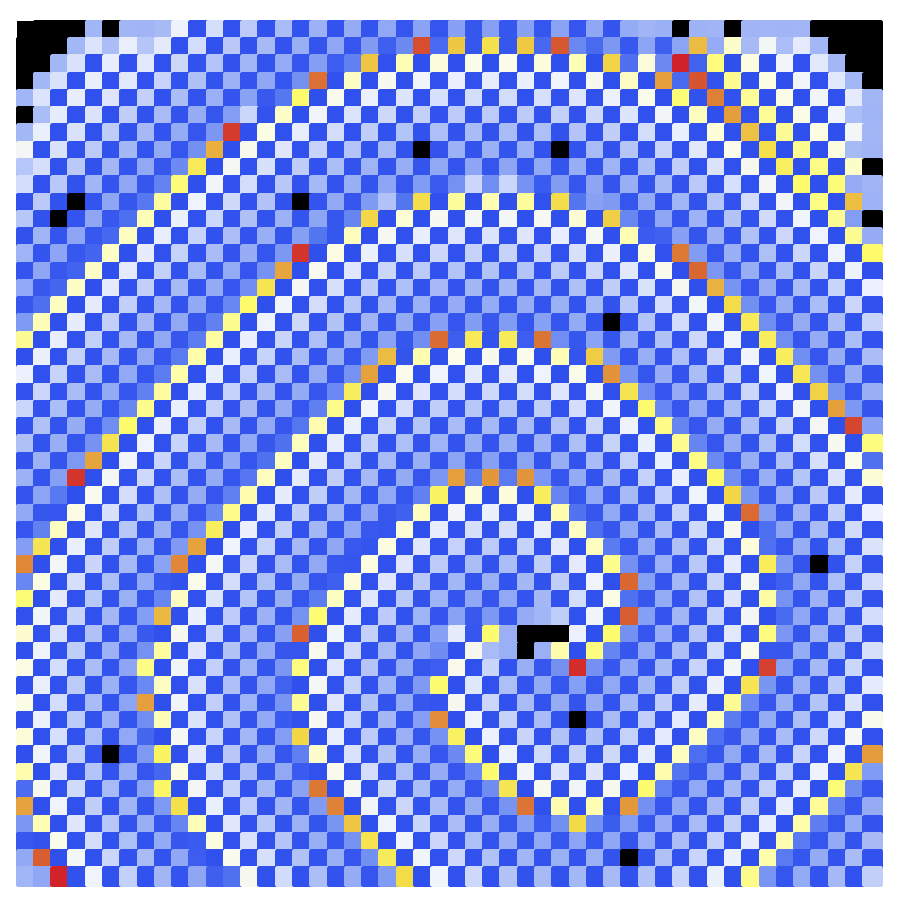}}\hfill
	\subfloat[]
		{\includegraphics*[width=0.23\columnwidth]{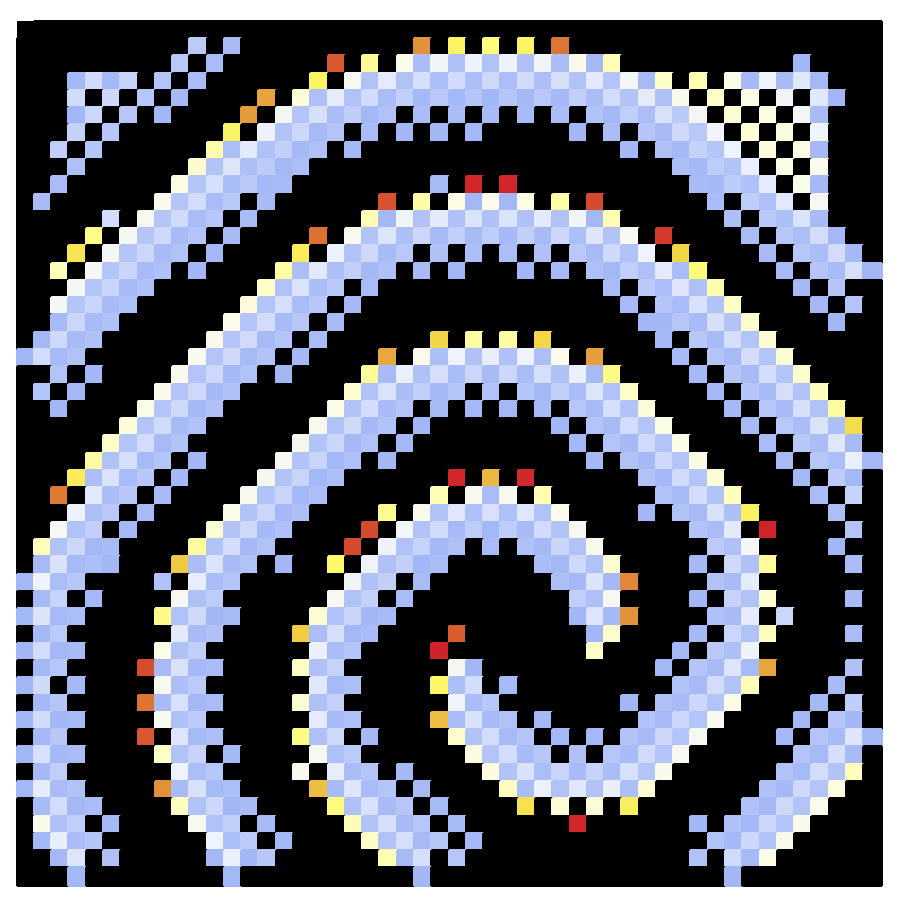}}\hfill
	\subfloat[]
		{\includegraphics*[width=0.23\columnwidth]{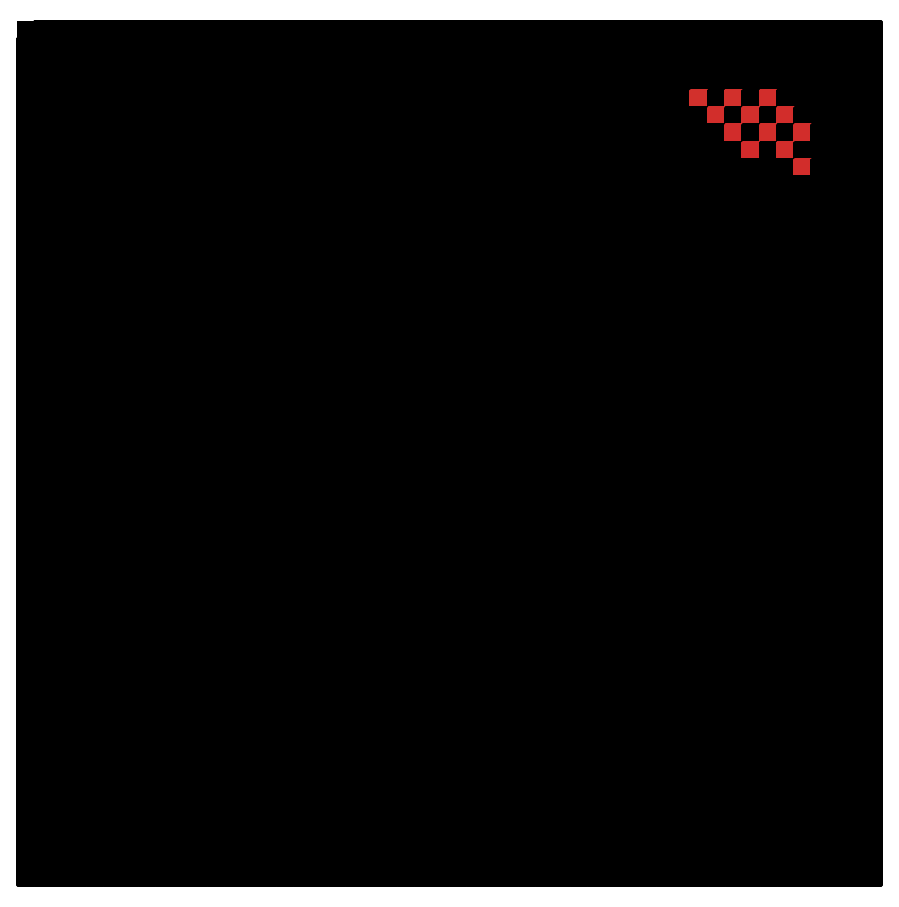}} \\
	\subfloat[]
		{\includegraphics*[width=1\columnwidth]{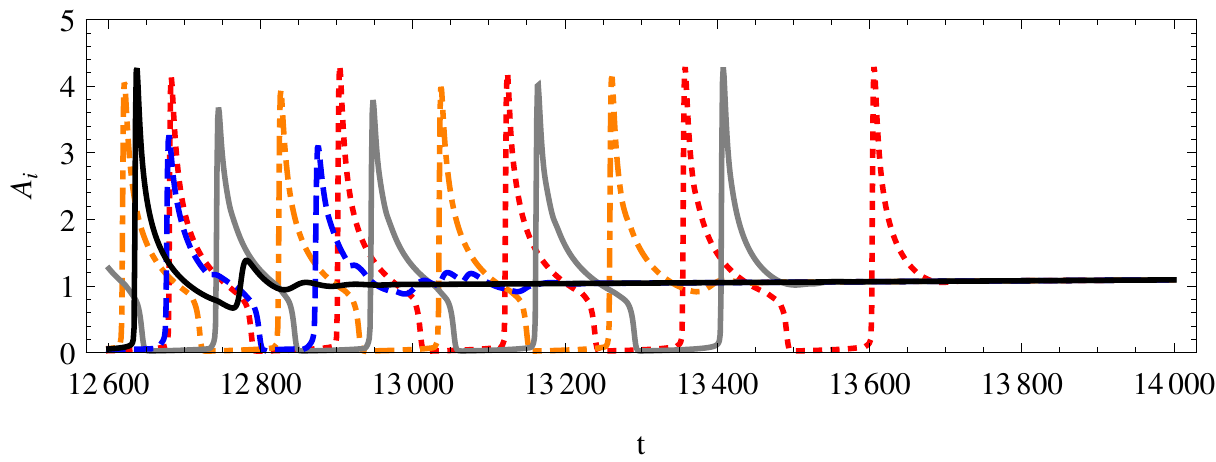}}
\caption{Conversion from a CO- to the upper CFP-regime on a SQU-UD lattice with f.b.c. and a gradual change of $\alpha$ at weak coupling. a) All units are in the oscillatory state at t=12600. b) The arrest starts at the upper corners and at the center of the spiral (black spots). The little crosses in b) are no indication for arrest, but for instants when the oscillator phases cross zero. c) Along the boundaries and the  black arms of the spiral the oscillations are arrested. d) The conversion is almost complete apart from an oscillating island (red spot). e) Time evolution of $A_{i,j}$ for a selection of five units to illustrate the subsequent arrest. For further details see the text.} \label{fig6:nucl}
\end{figure}

Let us first consider the case of fixed $\alpha$, chosen out of a small interval in the transition regime ($\alpha\in [99.5,100.0]$) of the SQU-UD and an even smaller interval for the SQU-OC ($\alpha\in [99.8,100.0]$). We then obtain transients up to $\sim 10000$ time units that are characterized by coexisting, individually oscillating and ``frozen" units. The behavior of individual units may change from first being oscillatory and then arrested or vice versa. Let us next consider a  gradual change of $\alpha$ over the regime $[99.5,110.0]$. For this case  we can pursue to a certain extent which oscillators get arrested first and how the arrest spreads over the lattice. Starting from a CO-regime (figure~\ref{fig6:nucl} a)), it is at the corners and the centers of the spirals (black spots in figure~\ref{fig6:nucl} b)), where the oscillators get arrested first, and the arrest then propagates along the boundaries and the black arms of the spirals (figure~\ref{fig6:nucl} c)), until a small island of oscillatory units is left in the bulk of fixed-point behavior (red spot in figure~\ref{fig6:nucl} d)) that afterwards gets arrested as well. The values of the arrested oscillator phases are not those of the rotating unit at the time instant of arrest, as it is assumed in a clock-wavefront mechanism, but of the fixed-point value at the given value of $\alpha$. This is seen from figure~\ref{fig6:nucl} e) which shows the time evolution of five A-values: the black curve corresponds to the upper left corner point $A_{1,1}$ which freezes first, the blue curve to the center of the spiral $A_{30,37}$, which freezes next; in our selection of ``freezing events" they are followed by a unit along the boundary $A_{25,1}$ (orange curve), where the freezing spreads, followed by a unit close to the center of the spiral $A_{28,33}$ (grey curve) and a unit somewhat in the bulk close to the upper right corner $A_{43,8}$ (red curve). The selection of the last three units is somewhat arbitrary. In principle it can be used for pursuing the chronological order of how the freezing spreads over the lattice. The time instants of the snapshots of figure~\ref{fig6:nucl} are 12600 (a), 13250 (b), 13500 (c) and 13710 (d). In particular panel c) illustrates the spreading along the arms of the spiral.

What distinguishes corners and the center of the spiral from other locations on the grid are the number of neighbors with a given phase value in the oscillatory regime. The less neighbors of a given unit can be  in a state that would be ``preferred" by this unit and stabilize its own state, the more sensitive it reacts upon a change of the parameter values, here of $\alpha$. Clearly it depends on the coupling sign and strength in competition with its individual dynamics, which neighboring states are favored  by the unit.
In systems of condensed matter it is the interface free energy  between liquid and gaseous phases in comparison to the free energy in the bulk that determines the conditions for the survival of a nucleus of liquid in the bulk of gas, or vice versa. The nucleus has to exceed a critical size to survive and grow. Here it remains to quantify the condition  for the ``conversion" of CO- into CFP-regimes in terms of basins of attraction. Concretely, it is the basin of attraction of the fixed-point state, first approached by the unit in the center of the spiral and at the corners, that should exceed a ``critical size" to attract the value of A, and arresting its oscillations, so that other oscillators will follow. We shall not pursue this question here, as a further zoom into the neighborhood of first ``freezing" units would be needed within the high-dimensional and rough attractor landscape.

\subsubsection{Influence of the coupling strength in the repressilator-limit}\label{secstrong}
Stronger repressive coupling accelerates the synchronization towards the pattern of the stationary state in the CO-regime.
If we compare movies for the SQU-OC-lattice (with p.b.c. and a gradual change of $\alpha$) for weak and strong coupling, it is for strong coupling that the units synchronize much faster  towards a state with two clusters of oscillators, that is a chessboard pattern of oscillator phases on the square lattices (cp. the movies \cite{mov:squ_oc_weak} and \cite{mov:squ_oc_strong}). Tuning $\alpha$ towards the second CFP-regime, the system evolves to a one-cluster fixed point solution, so that the individual dynamics obviously dominates the coupling term even for strong couplings.

\begin{figure}[ht]
\center
	\includegraphics*[width=1\columnwidth]{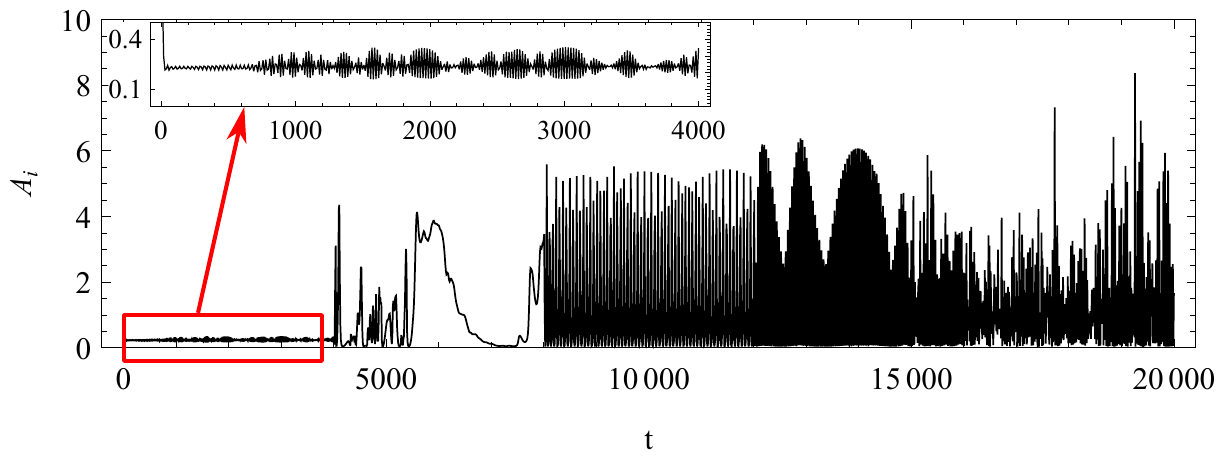}\\
	\includegraphics*[width=1\columnwidth]{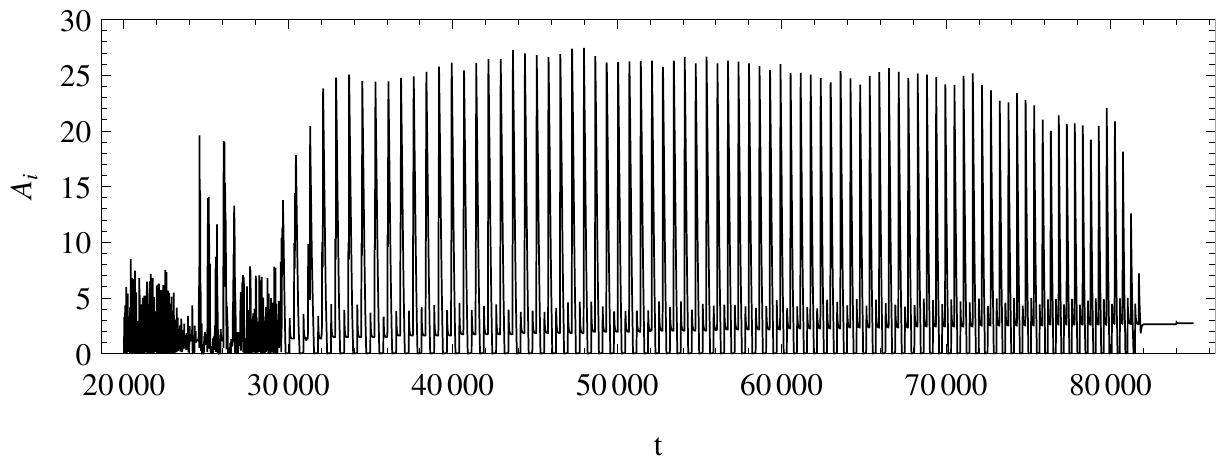}
\caption{Time evolution of a representative single unit, $A_{50,25}$,  for strong coupling $\beta_R=10.0$ on the HEX-OC-lattice, while $\alpha$ is slowly varied. The inset shows a zoom into the first 4000 t.u.. The first fixed-point regime is absent, and in the expected oscillatory regime we have intervals with chaotic behavior.}  \label{fig8:coupl}
\end{figure}

For strong coupling and the HEX-OC topology, the lower CFP-regime is completely absent and the former CO-regime becomes chaotic with intervals of oscillating behavior, as is best seen from the time evolution of oscillator phases, one of which is displayed in figure~\ref{fig8:coupl}. For $\alpha$ large enough, the system evolves to a fixed point, for the simulation displayed in figure~\ref{fig8:coupl} it happens at $\alpha = 400$ which is reached at time 80000 t.u.. For smaller values of $\alpha$, the system is in the limit, in which the coupling term dominates over the individual dynamics, a limit that we call repressilator-limit.

\begin{figure}[ht]
\center
\includegraphics*[width=1\columnwidth]{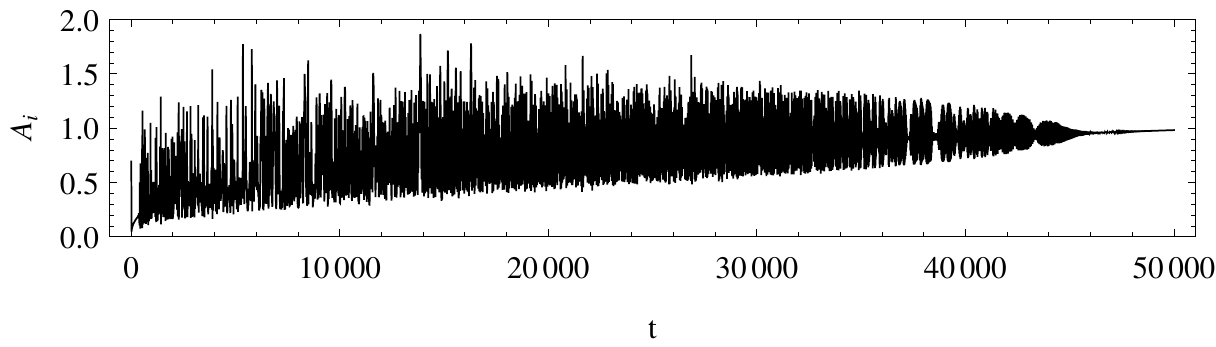}
\caption{Time evolution of a single representative unit $A_{50,25}$  on the HEX-OC-lattice, while $\alpha$ is fixed and $\beta_R$ is varied,  approaching  the repressilator-limit. After a small interval of $\beta_R\sim [0,5.25]$ of CFP-behavior during the first 350 time units, we see chaotic behavior for the next 35000 time units. When the coupling is strong enough, the system evolves to a second CFP-regime at stronger couplings. The parameters are $\alpha = 5.0$, $K=0.02$, $b=\gamma=0.01$, and $\Delta \beta_R = 0.015$, $\Delta t = 1$ t.u.} \label{fig9:repr}
\end{figure}

We are particularly interested in this limit by two reasons. On the experimental side no self-activating components have been identified yet in connection with the segmentation clock. From the theoretical perspective the individual dynamics of our BFUs is already rich and flexible, and the question arises as to whether a basic unit with a simpler individual dynamics would do it as well: when coupled to other such units, providing a simple mechanism for controlling the duration of oscillations along with a rich spectrum of transients, some of which may be relevant in the biological context as well. Therefore we studied our system in the limit of very strong coupling, where the effect of self-activation should be suppressed and the HEX-OC lattice resembles  a set of repressilators \cite{jensen2009}. Self-activating components are completely absent in repressilators. The role of the bifurcation parameter there is played by the repressive coupling, and a CO-regime exists just because of the coupling. Varying now in our system the coupling (not $\alpha$) from weak to strong on a HEX-OC lattice for small, but fixed $\alpha$ to realize the repressilator-limit, we see first a CFP-regime for $\beta_R\in\approx [0,6]$ (weak coupling), followed by chaotic behavior for intermediate coupling strengths, and again CFP-behavior for larger couplings, see figure~\ref{fig9:repr}. For larger lattices, the second CFP-behavior  happens only for very strong coupling ($\sim \beta_R = 720$ for a $50\times 50 $ lattice). This means, in the limit where we can neglect the effect of self-activation, we see chaotic or fixed-point behavior. Our results are in agreement with the results of \cite{jensen2009}. The other lattice topologies do not resemble the repressilator-lattice with three repressing units coupled in an elementary loop.  Therefore the positive feedback components in our system of coupled BFUs seem to be an essential ingredient for the simple control mechanism to work.

\subsubsection{Influence of the variation speed and boundary conditions}
\begin{figure}[ht]
\captionsetup[subfigure]{labelformat=simple}
\centering
	\subfloat[]
		{\includegraphics*[width=.230\columnwidth]{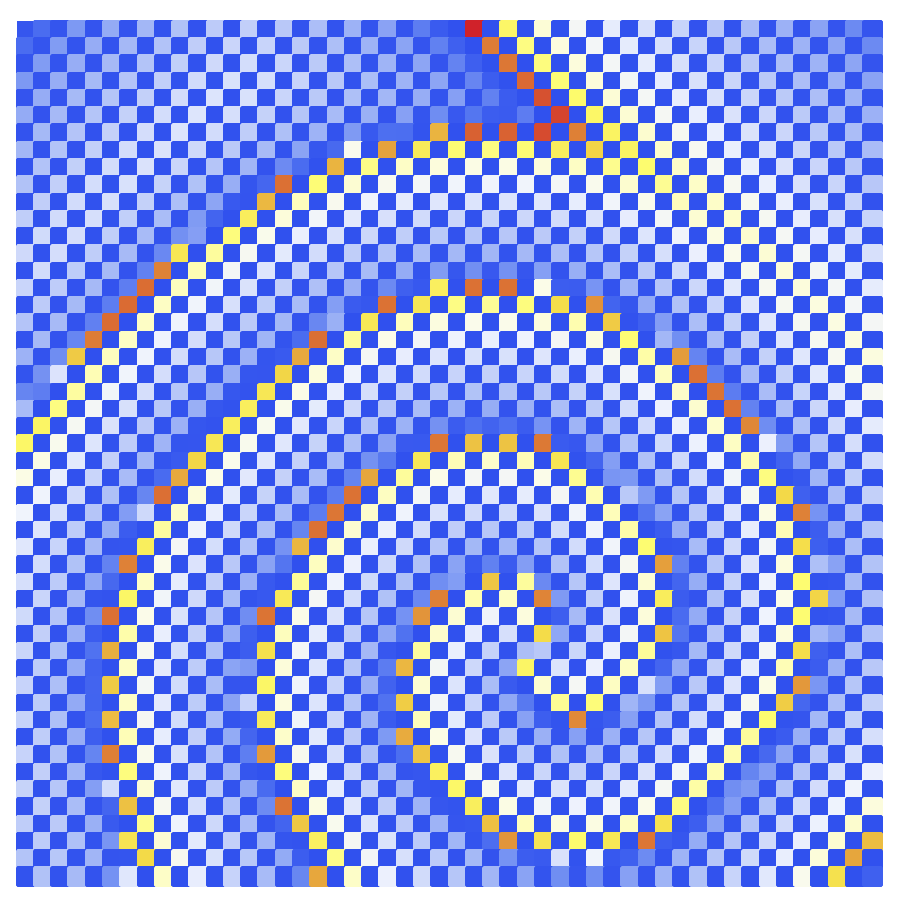}}\hfill
	\subfloat[]
		{\includegraphics*[width=.230\columnwidth]{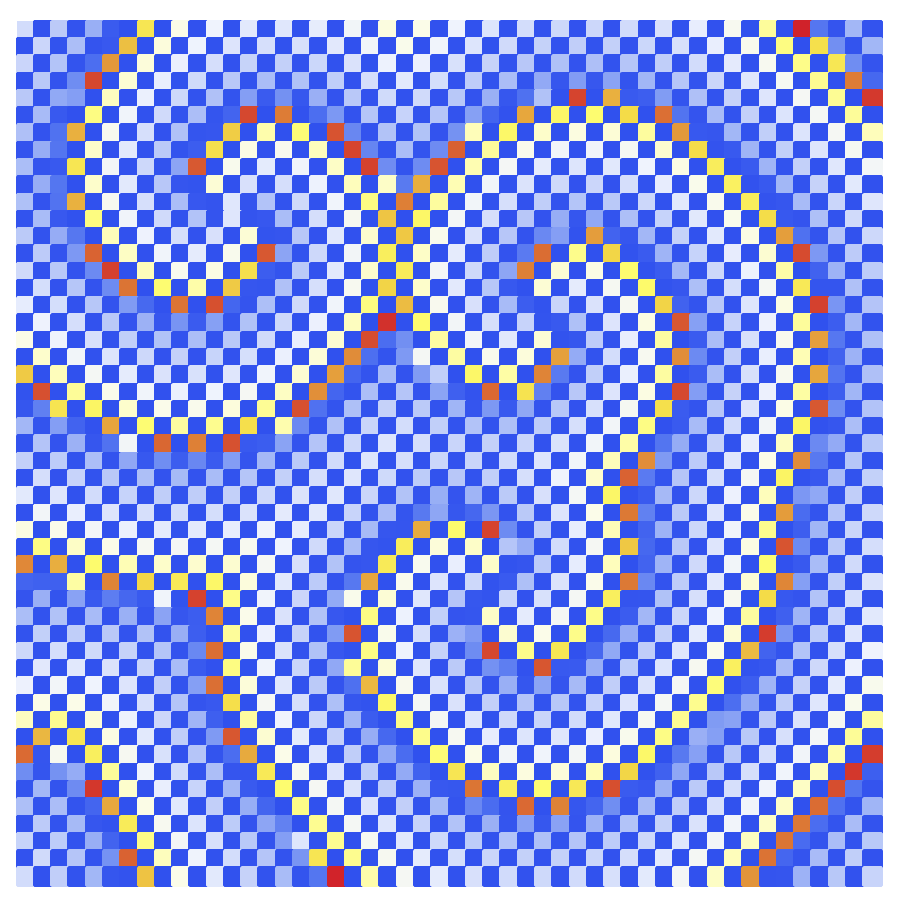}}\hfill
	\subfloat[]
		{\includegraphics*[width=.230\columnwidth]{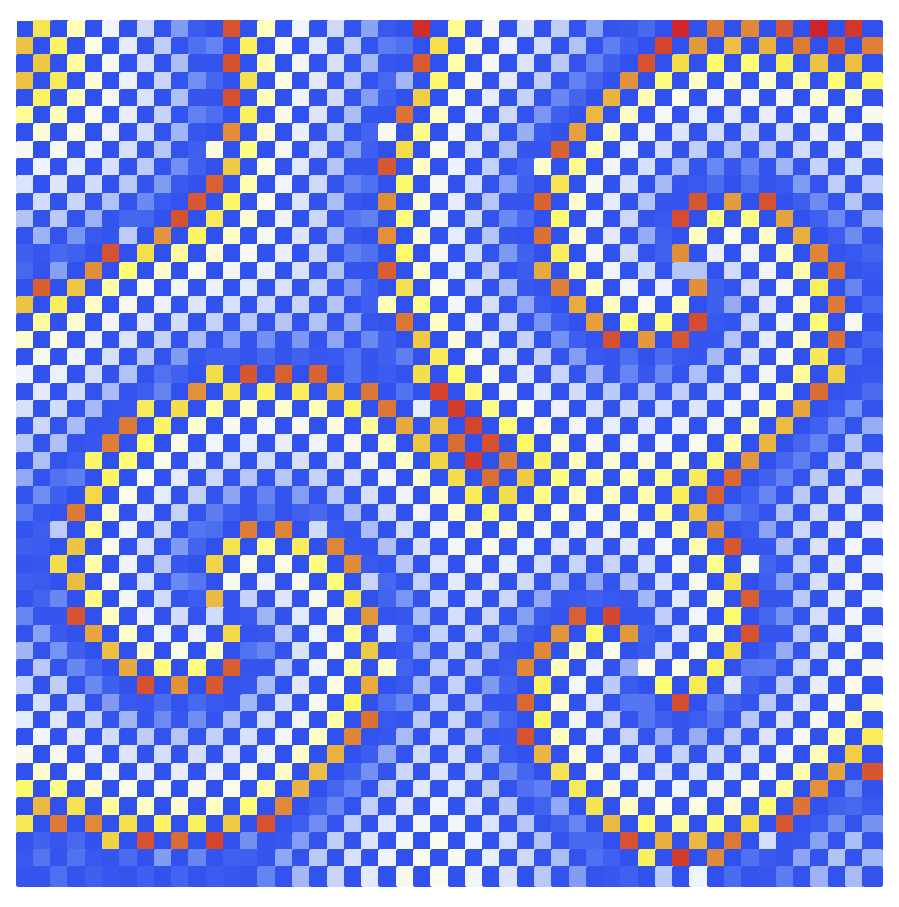}}\hfill
	\subfloat[]
		{\includegraphics*[width=.230\columnwidth]{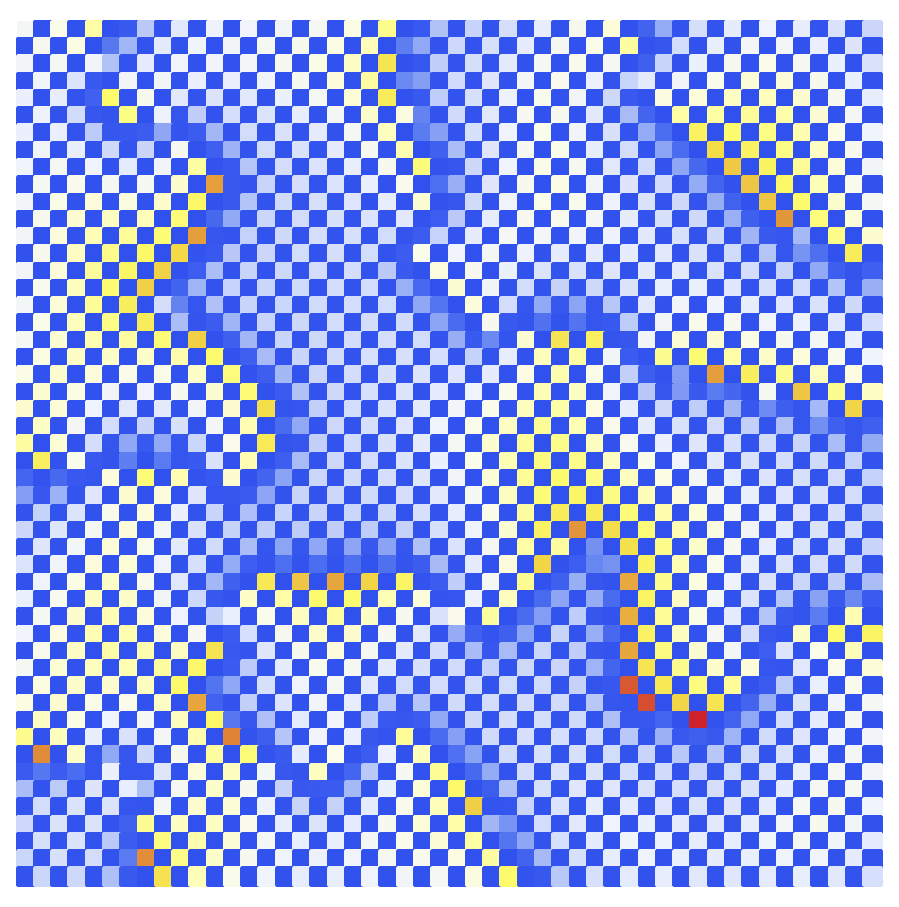}}
\caption{Influence of the initial conditions (a) and b)) and the speed (c) and d)) of varying $\alpha$ on a SQU-UD-lattice for weak coupling. Panels a) and b) show snapshots at $\alpha = 95.0$ and $95.5$, respectively, resulting from the same gradual change of $\alpha$, with different initial values of $\alpha_0 = 5.0$ and $1.0$, respectively. Panel c) and d) show snapshots at $\alpha = 95$ for slow and fast variation of $\alpha$, respectively. For both cases the initial value is $\alpha_0 = 5.0$, for d) spirals had no time to evolve.} \label{fig7:speed}
\end{figure}

In view of the control mechanism we mainly focused on a gradual change of $\alpha$, although realized in an non-adiabatic way. The slow and fast speeds correspond to iterated quenches with short (fast speed) or longer (slow speed) time intervals of constant $\alpha$ in between. The speed in these cases mainly influences the dynamics via the initial conditions for the subsequent value of $\alpha$, provided by the concentration values at the end of the preceding interval of fixed $\alpha$. In the case of multistability  the subsequent states sensitively depend on the initial conditions.
Figure~\ref{fig7:speed}  a) and b) illustrate the dependence on the initial conditions for otherwise the same gradual speed, where a different number of spirals is generated, for the case of a SQU-UD-lattice. For a fast variation of $\alpha$, where we change $\alpha$ every 100 t.u., which is of the same order as the phases period, the system has not enough time to form distinguishable spirals, cf. figure~\ref{fig7:speed} d) in contrast to c). In general, for a fast change of $\alpha$, some patterns have not enough time to evolve at all.

\begin{figure}[ht]
\captionsetup[subfigure]{labelformat=simple}
\centering
	\subfloat[]
		{\includegraphics*[width=0.18\columnwidth]{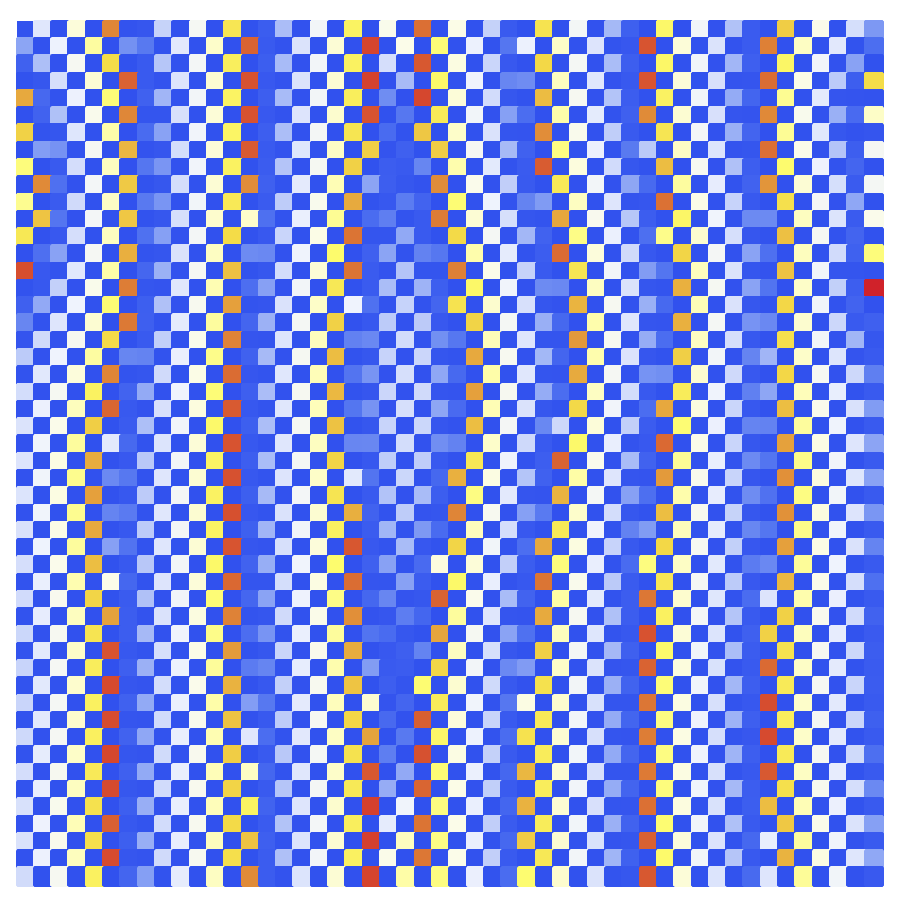}}\hfill
	\subfloat[]
		{\includegraphics*[width=0.18\columnwidth]{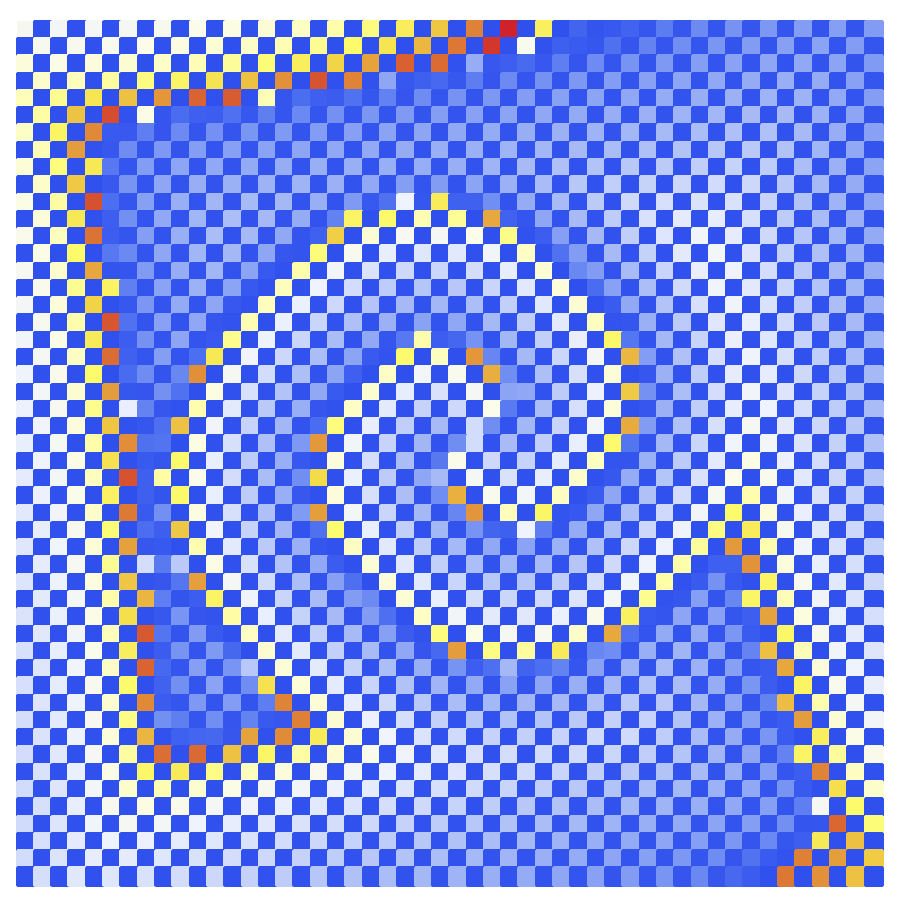}}\hfill
	\subfloat[]
		{\includegraphics*[width=0.18\columnwidth]{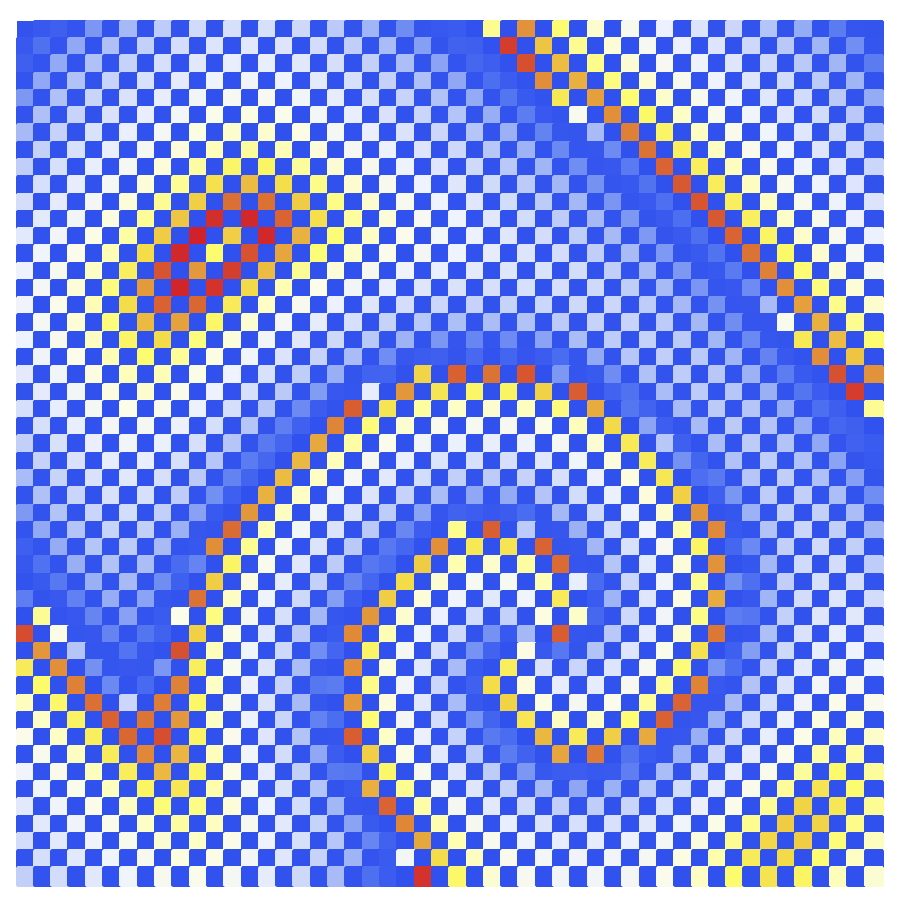}}\hfill
	\subfloat[]
		{\includegraphics*[width=0.18\columnwidth]{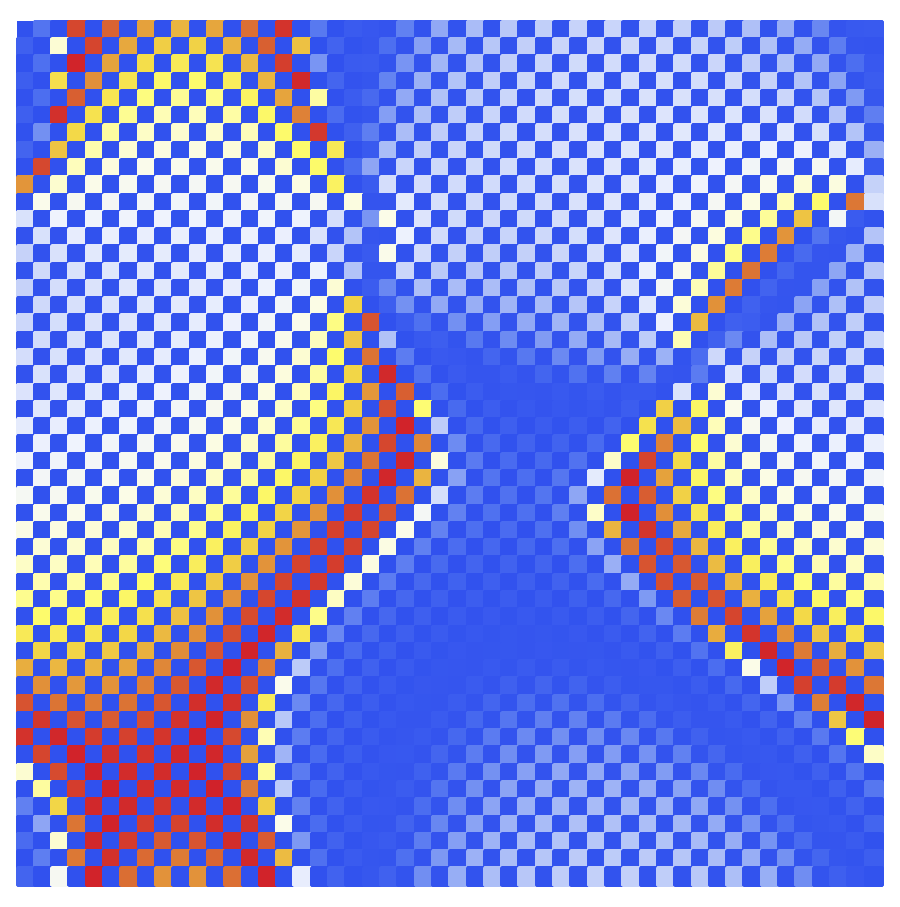}}\hfill
	\subfloat[]
		{\includegraphics*[width=0.18\columnwidth]{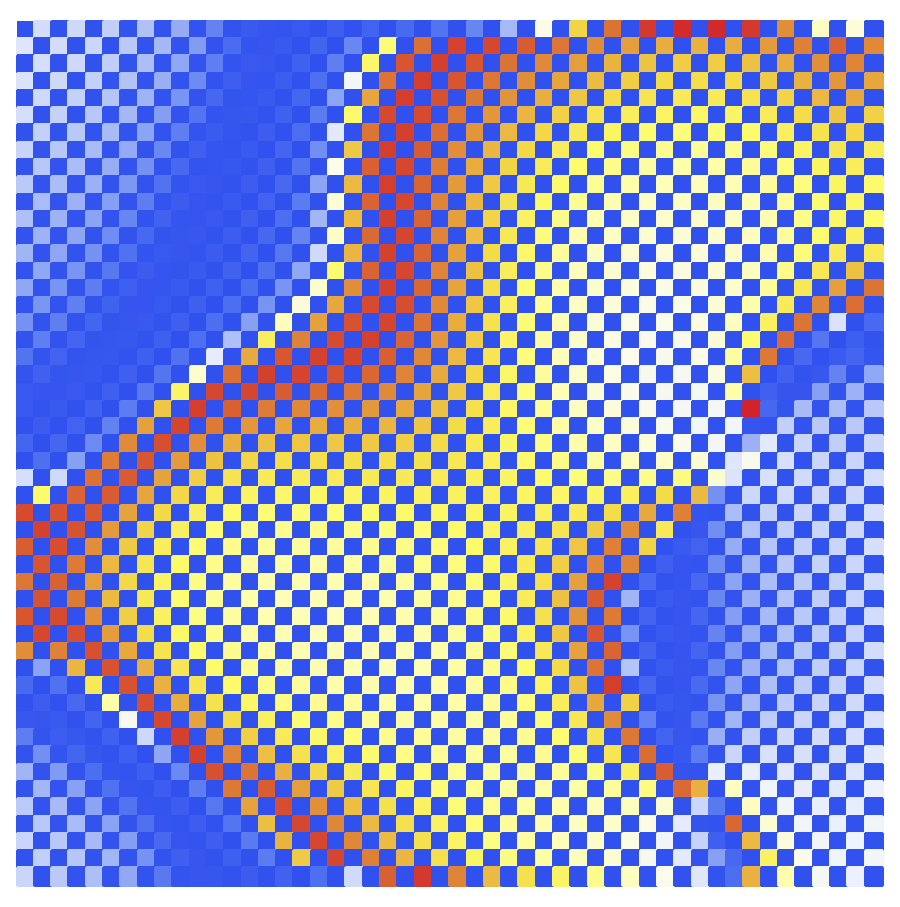}}
\caption{Influence of the boundary conditions. The first three panels a)-c) show snapshots of a SQU-SC-lattice for p.b.c., a SQU-UD lattice with f.b.c., and again a SQU-SC lattice with f.b.c., respectively. Panels d) and e) show the SQU-OC-lattice with p.b.c. and f.b.c., respectively. All snapshots are obtained for gradual variation of $\alpha$ and weak coupling.} \label{fig10:bc}
\end{figure}

The influence of p.b.c. seems to be more restrictive for lattices with plaquettes of the same chirality (SQU-SC figure~\ref{fig10:bc} a) and c)) with p.b.c. and f.b.c., respectively, or with no chirality (SQU-UD figure~\ref{fig10:bc} b)) than for lattices with opposite chirality like the SQU-OC-lattice with p.b.c. (d) and f.b.c. (e). In figure~\ref{fig10:bc} a) the spirals are replaced by stripes which propagate from the boundaries and delete the spirals when coalescing with them in the bulk. Obviously the very choice of boundary conditions may have a selective effect on the evolving transient patterns that must not be ignored in view of applications to biological systems. From simulations of larger lattices of size $100\times 100$ we conclude that their influence remains pronounced for larger sizes.

\subsubsection{A quasi one-dimensional arrangement}
\begin{figure}[ht]
\captionsetup[subfigure]{labelformat=simple}
\centering
	\subfloat[]
		{\includegraphics*[width=0.3\columnwidth]{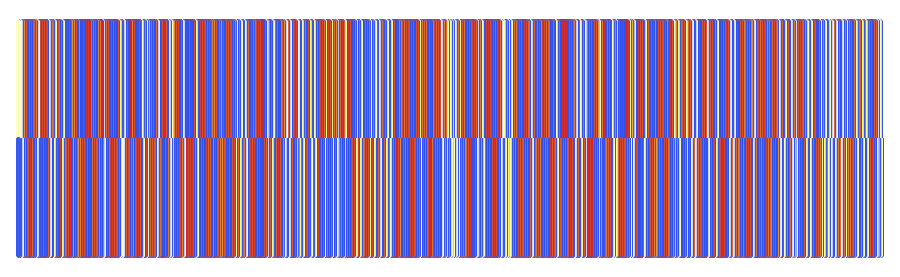}}\hfill
	\subfloat[]
		{\includegraphics*[width=0.3\columnwidth]{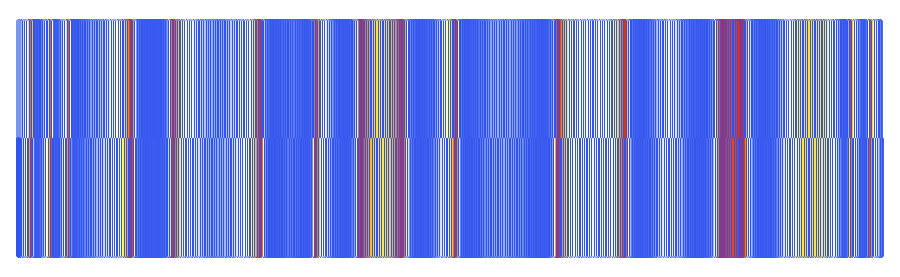}}\hfill
	\subfloat[]
		{\includegraphics*[width=0.3\columnwidth]{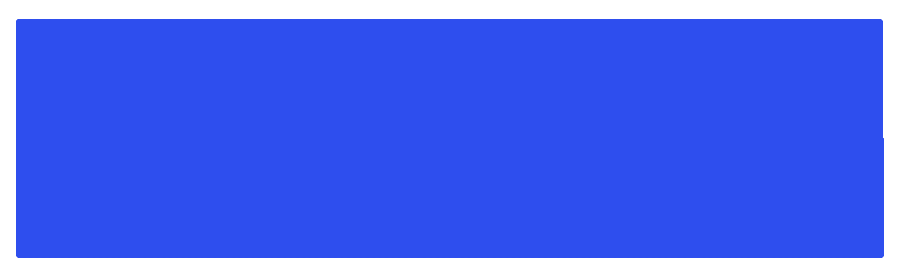}}
\caption{Stripes on a quasi one-dimensional ($1000 \times 2$) SQU-UD-lattice with f.b.c.. Panels a) and c) show fixed point regimes, b) the oscillatory regime. Snapshots are taken at times 50, 5000, and 15000 t.u. respectively. Other parameters are $\beta_R = 0.1$, $K=0.02$, $b=\gamma=0.01$, $\Delta \alpha = 0.0075$, $\Delta t = 1$ t.u..} \label{fig11:1D}
\end{figure}

In view of our original motivation in connection with the segmentation clock we also checked the type of patterns which evolve for a quasi one-dimensional topology chosen as a $1000 \times 2$ SQU-UD lattice with f.b.c.. Figure~\ref{fig11:1D}a) shows a multi-cluster fixed point, b) a multi-cluster oscillatory state with stripes that are traveling from the boundaries to the center of the lattice, and c) a one-cluster fixed point for large $\alpha$. It should be noticed that the mechanism of generating the stripes here is different from the clock and wavefront mechanism of \cite{1976} that is usually supposed to generate the stripes in connection with the segmentation clock.

\section{Summary and Conclusions} Coming back to our original motivation from the segmentation clock, it is two bifurcations in our system of coupled BFUs that initiate and arrest the oscillations when a single bifurcation parameter in monotonically increased. The speed of its variation determines the duration of the overall oscillation period and of the type of transient patterns. There is no need for finetuning the frequency profile via a space-dependent choice of input parameters, neither is there a need for an external field to induce an arrest of the oscillations, or for a higher-level counting mechanism implemented in the elongating embryo. (For a careful and detailed discussion of alternative mechanisms in connection with the segmentation clock we refer to \cite{count}.) Our mechanism of pattern generation is therefore different from the clock and wavefront mechanism \cite{1976}. We can create periodic patterns in space due to the traveling waves in the oscillatory period or as a collective  multi-cluster fixed point solution of the BFUs in the small-$\alpha$-regime, characterized by coexisting different but frozen values of the concentrations $A$ and $B$. Due to the repulsive coupling no neighboring oscillators share the same value unless the individual dynamics dominates the coupling term. Therefore the question arises whether there is any experimental support for our type of mechanism being at work in the segmentation clock. Given a system of coupled BFUs, each composed of a negative and positive feedback loop, the driving force for the onset and arrest of oscillations is the variation of a single bifurcation parameter. On the experimental side, retinoic acid (a morphogene) is an example of a molecule that increases in concentration monotonically when cells traverse the presomitic mesoderm from posterior to anterior, although the influence of this morphogene on oscillations is currently not (yet) understood on a molecular level and its relation to our parameter $\alpha$ is therefore open.

The second question concerns the realization of coupled BFUs in the segmentation clock. So far, negative feedback loops were identified in the zebrafish, where certain dimers repress the genes of their components \cite{schroeter2012}, but no indications for self-activating loops have been found so far. This does, of course, not exclude their existence. On the modeling side, for comparison we simulated both, a system of coupled repressilators, and our system of coupled BFUs in the limit of very strong coupling. As outlined above, the control of the duration of oscillations failed in the sense that the first and third fixed-point regime may be  absent or the oscillatory regime is replaced by a chaotic one. So our mechanism does not seem to work with these simpler building blocks.

From a more general perspective  the role of positive feedback loops in combination with negative ones was studied in \cite{tsai}. There it was shown that in general a combination of interlinked positive and negative feedback loops of interacting genes allows to adjust the frequency of a negative feedback oscillator while keeping the amplitude of oscillations nearly constant. At the same time, the combination with a positive feedback loop makes the system more robust against perturbations, and therefore improves its reliability. Although it seems to be open whether these observations are relevant for the segmentation clock, these results together with our observations may trigger a search for self-activating loops on the experimental side of the segmentation clock.

From the mere physics' perspective our system of coupled BFUs provides a rich repertoire of transient patterns, in particular of self-organized  pacemakers, and multiple inherent dynamically generated  time scales with a variety of attractors.  A closer zoom into the transition region, where the ``conversion" between collective oscillatory and collective fixed-point behavior happens, poses a challenge for further research.

\section*{Acknowledgments}
We would like to thank F. J\"ulicher and D. J\"org (Max-Planck Institute for the Physics of Complex Systems, Dresden) and L.G. Morelli (Universidad de Buenos Aires) for useful discussions about the segmentation clock. Financial support from the Deutsche Forschungsgemeinschaft (DFG, contract ME-1332/19-1) is gratefully acknowledged.

\end{document}